\documentclass[aps,prl,twocolumn,superscriptaddress]{revtex4-2}
\usepackage{graphicx,amssymb,amsmath,amsfonts,empheq}
\usepackage[dvipsnames]{xcolor}
\usepackage{color}
\usepackage{braket}
\usepackage{verbatim}
\usepackage{latexsym}
\usepackage{upgreek}
\usepackage{dsfont}
\usepackage{bm}
\usepackage{multirow}
\usepackage{enumitem}
\usepackage[normalem]{ulem}
\usepackage{url}
\usepackage{hyperref}
\usepackage{soul}
\hypersetup{
colorlinks = true,
linkcolor = [rgb]{0,0,0}, 
citecolor = [rgb]{0.13,0.55,0.13},
urlcolor = [rgb]{0.25, 0.41, 0.88}}
\usepackage{bbm}

\usepackage{tikz}
\usepackage{tikz-cd}
\usetikzlibrary{arrows}
\usetikzlibrary{intersections}
\usetikzlibrary{shapes.geometric}
\usetikzlibrary{decorations.pathmorphing, patterns,shapes}
\usetikzlibrary{decorations.markings}

\tikzset{
	partial ellipse/.style args={#1:#2:#3}{
		insert path={+ (#1:#3) arc (#1:#2:#3)}
	}
}

\tikzset{
	mid arrow/.style={postaction={decorate,decoration={
				markings,
				mark=at position .575 with {\arrow[#1]{stealth}}
	}}},
	near arrow/.style={postaction={decorate,decoration={
				markings,
				mark=at position .275 with {\arrow[#1]{stealth}}
	}}},
	far arrow/.style={postaction={decorate,decoration={
				markings,
				mark=at position .800 with {\arrow[#1]{stealth}}
	}}},
}

\pgfdeclarepatternformonly{south west lines}{\pgfqpoint{-0pt}{-0pt}}{\pgfqpoint{3pt}{3pt}}{\pgfqpoint{3pt}{3pt}}{
	\pgfsetlinewidth{0.4pt}
	\pgfpathmoveto{\pgfqpoint{0pt}{0pt}}
	\pgfpathlineto{\pgfqpoint{3pt}{3pt}}
	\pgfpathmoveto{\pgfqpoint{2.8pt}{-.2pt}}
	\pgfpathlineto{\pgfqpoint{3.2pt}{.2pt}}
	\pgfpathmoveto{\pgfqpoint{-.2pt}{2.8pt}}
	\pgfpathlineto{\pgfqpoint{.2pt}{3.2pt}}
	\pgfusepath{stroke}}

\renewcommand{\bar}{\overline}
\renewcommand{\i}{\mathrm{i}}
\renewcommand{\leq}{\leqslant}

\renewcommand{\Re}{\operatorname{Re}}

\newcommand{\Tr}{\operatorname{Tr}}

\newcommand{\bbR}{\mathbb{R}}

\newcommand{\calC}{\mathcal{C}}

\newcommand{\calO}{\mathcal{O}}

\newcommand{\calR}{\mathcal{R}}

\newcommand{\calT}{\mathcal{T}}

\newcommand{\eqnref}[1]{Eq.~\eqref{#1}}
\newcommand{\figref}[1]{Fig.~\ref{#1}}
\newcommand{\appref}[1]{Appendix.~\ref{#1}}

\definecolor{Acolor}{RGB}{242,120,121}
\definecolor{Bcolor}{RGB}{130,230,130}
\definecolor{Ccolor}{RGB}{153,163,252}

\begin{document}

\title{Extracting the Quantum Hall Conductance from a Single Bulk Wavefunction}

\author{Ruihua Fan} 
\author{Rahul Sahay} 
\author{Ashvin Vishwanath} 
\affiliation{Department of Physics, Harvard University, Cambridge, MA 02138, USA}

\begin{abstract}

We propose a new formula that extracts the quantum Hall conductance from a single (2+1)D gapped wavefunction. The formula applies to general many-body systems that conserve particle number, and is based on the concept of modular flow: i.e., unitary dynamics generated from the entanglement structure of the wavefunction.
The formula is shown to satisfy all formal properties of the Hall conductance: it is odd under time-reversal and reflection, even under charge conjugation, universal and topologically rigid in the thermodynamic limit.
Further evidence for relating the formula to the Hall conductance is obtained from conformal field theory arguments.
Finally, we numerically check the formula by applying it to a non-interacting Chern band where excellent agreement is obtained.  
\end{abstract}

\maketitle	

Since the discovery of the quantum Hall effect, the investigation of gapped topological phases has been at the forefront of quantum many-body physics~\cite{Qi:2010qag,Chiu:2015mfr,xiaogangReview}.
Such phases cannot be characterized by any local order parameters and all of their unusual properties emerge from the entanglement of the ground state wavefunction~\cite{XiaoGangQIBook}.
Quantifying this connection has been a persistent and interesting challenge for the past few decades.

For example, entanglement entropy in a ground state wavefunction reveals the total quantum dimension of anyons ~\cite{Hamma:2004vdz,LevinWen2005,Kitaev:2005dm} while access to multiple ground states on a torus allows for a determination of modular matrices characterizing the anyons \cite{Zhang:2011jd, VidalCincio}. Further information, such as the entanglement spectrum, reveals the protected edge states 
~\cite{LevinWen2005,Kitaev:2005dm,HaldaneLi,Fidkowski2009,Ari2009,PollmannEntSPT,Swingle:2011hu,Chandran2011,QiLudwig}. 
Yet, \emph{how to extract the quantum Hall conductance, the first discovered topological quantity, from the entanglement pattern of a gapped wavefunction alone} remains an open question. 
Here, by gapped wavefunction, we mean the ground state wavefunction of a gapped, local Hamiltonian. Besides its foundational interest, such a formula could help unravel the entanglement structure of local quantum systems and thus characterize limits to the efficient representations of quantum states by tensor networks or stabilizers~\cite{Dubail:2013pda, Wahl:2013rha, KapustinFidkowski2018,Siva:2021cgo,Liu:2021ctk}.

There are several approaches to extracting Hall conductance that do not invoke entanglement.
For free fermion systems, there are the TKNN formula~\cite{Thouless:1982zz} and Fredholm index formula~\cite{Bellissard1994,Simon1990,Kitaev:2005hzj}. 
While the latter idea has been generalized to the interacting case, its formalism uses quasi-adiabatic evolution and is more convenient to work with if the actual Hamiltonian, rather than a single wavefunction, is known~\cite{Bachmann:2020unp,KapustinHall2020}.
We also note earlier works on extracting the many-body Chern number from a single wavefunction based on connections to topological quantum field theory and surgery~\cite{Shiozaki:2017ive,Dehghani:2020jls}. This quantity coincides with the  Hall conductance in free-fermion systems but is a distinct quantity in generic interacting systems (see \cite{Dehghani:2020jls} for details).

In this letter, we conjecture a new formula, with analytical and numerical support, which uses entanglement to compute the Hall conductance from a single wavefunction directly. The formula is proposed for generic interacting systems without assuming translational invariance~\footnote{The close relation between entanglement spectrum (ES) and protected edge states suggests a possible strategy based on the ES. However, typically, translation symmetry needs to be invoked to fully extract information present in the ES.}.

The Hall conductance relies on charge conservation, which the entanglement spectrum of a state is blind to~\footnote{One can in principle use the $U(1)$ resolved ES and study the spectral flow, which, however, requires a continuous family of states and is not convenient to work with.}.
Therefore, more data is necessary, motivating our use of the \emph{modular Hamiltonian} and \emph{modular flow} in the formula presented here.
In a recent advance, the modular Hamiltonian has appeared in a proposed formula for the chiral central charge~\cite{Kim:2021gjx,Kim:2021tse}, which can be viewed as entanglement transport under the modular flow \cite{Fan:2022ukm}. 
(See also \cite{Siva:2021cgo,Liu:2021ctk} for related works.)
Here we are interested in the Hall conductance $\sigma_{xy}$, a distinct topological invariant, which requires an additional conserved $U(1)$ charge $Q$.

\textit{Proposed Formula for Hall Conductance}--- 
Consider a lattice system on the plane with a $U(1)$ symmetry that is generated by the total charge operator $Q = \sum_i n_i$, where $i$ labels the lattice sites. 
Let $\ket{\psi}$ be a $U(1)$ symmetric gapped wavefunction, $\rho_D = \Tr_{\bar{D}} \ket{\psi} \bra{\psi}$ the reduced density matrix of any given region $D$. 
The modular Hamiltonian is defined as $K_D := -\ln \rho_D$, which is a Hermitian operator with a lower-bounded spectrum and conserves the total number of charges $Q_D$ in that region, i.e., $[Q_D,K_D]=0$. 
The modular Hamiltonian generates a unitary evolution $U_D(s) := e^{-\i sK_D}$, called modular flow, where the modular time $s$ is dimensionless~\cite{Haag:1992hx}.
The Hall conductance is encoded in a charge response under modular flow at the linear order in $s$.

Specifically, divide the plane into four regions $A,B,C$ and $D$, whose linear sizes are larger than the correlation length, and every three regions meet at a point once and only once~\cite{Kitaev:2005hzj,Kim:2021gjx}. 
Let $K_{\bullet}$ and $Q_{\bullet}$ denote the modular Hamiltonian and charge of the corresponding region.
The response function we focus on is:
\begin{equation}
\label{eq:sigmaxy formula 1}
\begin{gathered}
\includegraphics[width=2cm]{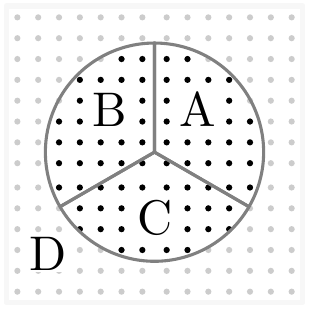}
\end{gathered}\quad
\begin{aligned}
	\Sigma(\psi;\, & A,B,C) \\
	&= \frac{\i}{2} \braket{\psi|[K_{AB},Q_{BC}^2]|\psi}\,,
\end{aligned}
\end{equation}
i.e., the charge response in $BC$ under the modular flow on $AB$. 
The modular Hamiltonian itself is non-universal and ultraviolet divergent, but the expectation value of the commutator will be argued to be topological and insensitive to details of the wavefunction.
Our proposal for Hall conductance $\sigma_{xy}$ is:
\begin{equation}
\label{eq:conjecture}
    \Sigma(\psi;A,B,C) = \sigma_{xy}\,.
\end{equation}
Note that the left-hand side containing $Q_{BC}^2$ is consistent with the fact that the unit of Hall conductance is $e^2/\hbar$.

The rest of the letter presents evidence for this formula from various angles. 
We first show that the formula shares the same general properties as the Hall conductance.
We then establish their quantitative connection via the bulk-edge correspondence and conformal field theory (CFT) arguments.
Our discussion, which uses the defect operator, provides a unifying viewpoint on the two different formulas for the chiral central charge and Hall conductance.
Finally, we provide numerical evidence using both lattice and continuum free-fermion models.

\emph{General properties}---
Any formula for the Hall conductance needs to meet the following minimal requirements. 
First of all, $\Sigma(\psi;A,B,C)$ should be real and additive under stacking, i.e., on combining independent systems:
\begin{equation}
\begin{aligned}
    \Sigma(\psi_1\otimes\psi_2;&A,B,C) = \\
    &\Sigma(\psi_1;A,B,C) + \Sigma(\psi_2;A,B,C)\,,
\end{aligned}
\end{equation}
which are clear from its definition.
Although additivity may seem trivial, other topological quantities e.g., the many-body Chern number do not possess this property.
We now show that $\Sigma(\psi;A,B,C)$ also satisfies other requirements that are not manifest in the construction.
It has the same $CRT$ transformation rules as the Hall conductance, namely, it (i) is even under charge conjugation:
\begin{equation}
\label{eq:charge conjugation}
    \calC: \quad \Sigma(\psi;A,B,C) = \Sigma(\calC\psi;A,B,C)\,,
\end{equation}
and (ii) is odd under the reflection
\begin{equation}
\label{eq:reflection}
    \calR: \quad \Sigma(\psi;A,B,C) = -\Sigma(\psi;B,A,C) \,,
\end{equation}
and (iii) is also odd under the time-reversal
\begin{equation}
\label{eq:time reversal}
    \calT: \quad \Sigma(\psi;A,B,C) = - \Sigma(\calT\psi; A,B,C)   \,.
\end{equation}
Furthermore, (iv) it is topological, in the sense that:
\begin{equation}
\label{eq:topological}
    \Sigma(\psi;A,B,C) = \Sigma(\psi;A',B',C')\,,
\end{equation}
where $A',\,B',\,C'$ are smoothly deformed subregions with the same topology as $A, \, B,\,C$.
Finally, (v) it is universal
\begin{equation}
\label{eq:deform Hamiltonian}
    \Sigma(\psi;A,B,C) = \Sigma(\psi';A,B,C)\,,
\end{equation}
where $\psi'$ is equal to $\psi$ deformed by a $U(1)$ symmetric local operator.

The justification of these properties uses three ingredients.
The first one is a conversion formula that follows the Schmidt decomposition
\begin{equation}
\label{eq:modular duality}
    K_D \ket{\psi} = K_{\bar D} \ket{\psi}\,,
\end{equation}
where $K_D$ and $K_{\bar D}$ are the modular Hamiltonians of the state $\ket{\psi}$ for an arbitrary region $D$ and the complement $\bar{D}$, respectively.
The second is the clustering property, the correlation function of two operators $O_x$ and $O_y$ factorizes when their distance $|x-y|$ is larger than the correlation length $\xi$
\begin{equation*}
    \braket{O_x O_y}_\psi = \braket{O_x}_\psi \braket{O_y}_\psi + \calO(e^{-|x-y|/\xi})\,,
\end{equation*}
where $\braket{\cdot}_\psi$ is the expectation value in the state $\ket{\psi}$.
The third one is an assumption on the decomposition of modular Hamiltonians when they act on the state:
\begin{equation*}
\begin{tikzpicture}[baseline={(current bounding box.center)}]
\draw (0,0) -- (2,0) -- (2,0.8) -- (0,0.8) -- (0,0);
\draw (0.65,0) -- (0.65,0.8);
\draw (2-0.65,0) -- (2-0.65,0.8);
\node at (0.325,0.4) {\scriptsize $A$};
\node at (0.975,0.4) {\scriptsize $B$};
\node at (2-0.325,0.4) {\scriptsize $C$};
\end{tikzpicture}\quad
\begin{aligned}
(K_{AB} & + K_{BC}) \ket{\psi}  \\
&\approx (K_{ABC} + K_B)\ket{\psi}\,,
\end{aligned}
\end{equation*}
where $A$ and $C$ do not meet directly~\cite{KitaevTalk}.
We expect the decomposition to hold when the linear sizes of the subregions are larger than the correlation length.
The proof of the $CRT$ transformation rules only uses \eqnref{eq:modular duality}. 
The argument of the topological rigidity and universal nature uses the other two ingredients as well. The third assumption was also used in \cite{Kim:2021gjx} to argue the rigidity of the modular commutator formula for the chiral central charge as well.

\paragraph{Charge conjugation} 
The charge conjugation sends the charge operator from $Q_D$ to $N_D - Q_D$, where $N_D$ is the maximal number charge of the region $D$. Here, we have assumed a finite Hilbert space per site. 
Thus, the right-hand side of \eqnref{eq:charge conjugation} is essentially
\begin{equation*}
\begin{aligned}
    \langle[K_{AB}, & (N_{BC} - Q_{BC})^2] \rangle_\psi \\
    &= 
    \braket{[K_{AB}, -2 N_{BC}Q_{BC} + Q_{BC}^2]}_\psi\,,
\end{aligned}
\end{equation*}
which can be shown to equal to the left-hand side by using $\braket{[K_{AB}, Q_{BC}]}_\psi = 0$ (see \appref{app:first law}). 

\paragraph{Reflection}
For a symmetric choice of $A,B$ and $C$, reflection effectively interchanges $A$ and $B$.
In fact, \eqnref{eq:reflection} holds more generally even when $A,B,C$ have unequal sizes and different shapes.
Nevertheless, we still call \eqnref{eq:reflection} the reflection transformation. Proving it is equivalent to showing the vanishing of the following quantity
$$
\begin{aligned}
    \langle[K_{AB},& Q_{BC}^2 + Q_{AC}^2] \rangle_\psi \\
    & =\langle[K_{AB}, Q_{BC}^2 + Q_{AC}^2 - Q_{ABC}^2]\rangle_\psi\,,
\end{aligned}
$$ 
where we have used $\braket{[K_{AB}, Q_{ABC}^2]}_\psi=0$ in the second step, i.e., modular flow in $AB$ does not change the charge in $ABC$.
Expanding the charge operators in terms of $Q_A,Q_B$ and $Q_C$ gives
$$
    \langle[K_{AB}, Q_{C}^2 - 2 Q_{A} Q_{B}]\rangle_\psi \stackrel{\text{Eq.}\eqref{eq:modular duality}}{=} 0,
$$
which finishes the proof. 

\paragraph{Time reversal}
The time reversal keeps the charge operator invariant. \eqnref{eq:time reversal} follows from $\braket{\calT\psi| \calT \phi} = \braket{\phi | \psi}$ and the hermiticity of $K_{AB}$ and $Q_{BC}$.

\begin{figure}
\begin{tikzpicture}[scale=0.9,baseline={(current bounding box.center)}]
	\draw[dashed, gray] (-1.5,-1.5) -- (-1.5,1.5) -- (1.5,1.5) -- (1.5,-1.5) -- (-1.5,-1.5);
    \filldraw[fill=red!10] (-0.7,0.65) circle (0.3);
    \node at (-0.8,0.75) {\scriptsize $b$};
    \filldraw[fill=white] (0,0) circle (1);
    \draw (0,0) -- (0,1);
	\draw (0,0) -- (0.85,-0.5);
	\draw (0,0) -- (-0.85,-0.5);
	\node at (1.1,-1.2) {\scriptsize $D$};
	\node at (0.4,0.4) {\scriptsize $A$};
	\node at (-0.4,0.4) {\scriptsize $B$};
	\node at (0,-0.55) {\scriptsize $C$};
	\node at (-1.2,1.2) {\scriptsize (a)};
\end{tikzpicture}\hspace{15pt}
\begin{tikzpicture}[scale=0.9,baseline={(current bounding box.center)}]
	\draw[dashed, gray] (-1.5,-1.5) -- (-1.5,1.5) -- (1.5,1.5) -- (1.5,-1.5) -- (-1.5,-1.5);
    \filldraw[fill=red!10] (-0.8,-0.6) circle (0.3);
    \node at (-0.9,-0.7) {\scriptsize $b$};
    \filldraw[fill=white] (0,0) circle (1);
    \draw (0,0) -- (0,1);
	\draw (0,0) -- (0.85,-0.5);
	\draw (0,0) -- (-0.85,-0.5);
	\node at (1.1,-1.2) {\scriptsize $D$};
	\node at (0.4,0.4) {\scriptsize $A$};
	\node at (-0.4,0.4) {\scriptsize $B$};
	\node at (0,-0.55) {\scriptsize $C$};
	\node at (-1.2,1.2) {\scriptsize (b)};
\end{tikzpicture}
\caption{Geometric deformation. Undeformed regions A,B,C with: (a) Deformation "b" is away from the triple contact points. (b) Deformation is close to one triple contact point.}
\label{fig:geometric deformation}
\end{figure}
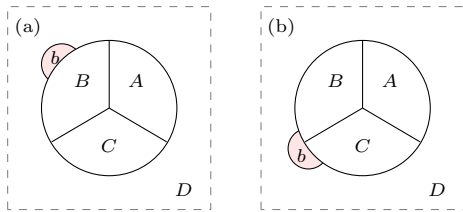

\paragraph{Topological}
We first deform the boundary between regions far from any triple contact point.
One typical example is \figref{fig:geometric deformation}~(a), where a small blob $b$ is removed from $D$ and attached to $B$. We write the change of $\Sigma$ as the sum of three terms
\begin{equation*}
    \begin{aligned}
    2\i\Delta\Sigma=& \braket{[K_{ABb}, Q_b^2]}_\psi \\
    & + \braket{[K_{ABb}, 2Q_b Q_{BC}]}_\psi \\
    & + \braket{[K_{ABb} - K_{AB}, Q_{BC}^2]}_\psi\,,
    \end{aligned}
\end{equation*}
and show that each vanishes individually.
We use \eqnref{eq:modular duality} to replace $K_{ABb}$ with $K_{CD\setminus b}$, the modular Hamiltonian on the complement region.
The vanishing of the first term becomes clear.
The vanishing of the second term requires the clustering property, i.e., 
$$
\begin{aligned}
    \braket{[K_{CD\setminus b},Q_b Q_{BC}]}_\psi
    \approx  \braket{Q_b}_\psi \braket{[K_{CD\setminus b}, Q_{BC}]}_\psi = 0
\end{aligned}
$$
where we used that the support of $[K_{CD\setminus b}, Q_{BC}]$ is far from the blob $b$.
The vanishing of the third term follows from the decomposition $K_{CD\setminus b} \rightarrow K_{CD} + K_{D\setminus b} - K_D$, where we have to assume that the blob $b$ is larger than the correlation length.

Next, consider deformations that are close to one of the triple contact points. One typical example is shown in \figref{fig:geometric deformation}~(b). 
We write the change of $\Sigma$ as
\begin{equation*}
\begin{aligned}
    2\i\Delta\Sigma = & \langle [K_{ABb}, Q_{BbC}^2] \rangle - \langle [K_{AB}, Q_{BC}^2] \rangle \\
    =& \langle [K_{ABb}, Q_{BbC}^2 - Q_{ABbC}^2] \rangle - \langle [K_{AB}, Q_{BC}^2] \rangle
\end{aligned}
\end{equation*}
where we imitate the proof of reflection transformation to add $Q_{ABbC}^2$ in the second step. The vanishing of the above expression then follows the decomposition $K_{ABb} \rightarrow K_{AB}+K_{Bb} - K_B$.

\paragraph{Universal} We deform the state $\ket{\psi}$ to $\ket{\psi_x'} \propto \ket\psi + O_x \ket\psi$ by a $U(1)$ symmetric local operator at position $x$, and show
\eqnref{eq:deform Hamiltonian}.
If $O_x$ is inside a region and far from any boundary, the clustering property implies that the reduced density matrices and thus modular Hamiltonians of other regions do not change.
One can then prove \eqnref{eq:deform Hamiltonian} by using \eqnref{eq:modular duality} and the clustering property repetitively (see \appref{app:rigidity}).
The case where $O_x$ is close to a boundary can be justified by exploiting the topological rigidity of $\Sigma(\psi;A,B,C)$, similar to the argument in \cite{Kitaev:2005dm}.

Although the results thus far are compatible with identifying $\Sigma \propto \sigma_{xy}$ in \eqnref{eq:conjecture}, the constant of proportionality must be determined. In particular we need to eliminate the trivial possibility that \eqref{eq:conjecture} always vanishes. We turn to this task next.

\emph{Connection to Hall conductance}---
The universal property \eqnref{eq:deform Hamiltonian} allows us to deform the state to a nicer form. 
Therefore, without loss of generality, we can consider gapped states whose edge can be described by a (1+1)D conformal field theory (CFT) with the central charge $c$ and $\bar{c}$. 
The modular Hamiltonian of a simply connected region $D$ with a smooth boundary can be approximated by the same CFT Hamiltonian supported on the boundary $\partial D$~\cite{Kitaev:2005dm,HaldaneLi}
\begin{equation}
\label{eq:KD of 2d bulk}
    \rho_D = \frac{1}{Z_D} e^{-K_D}\,,\quad
    K_D = \beta H_{CFT}\,,
\end{equation}
where $\beta$ is a non-universal constant. $H_{CFT}$ is defined by the Virasoro generators $L_0$ and $\tilde{L}_0$
\begin{equation*}
    H_{CFT} = \frac{2\pi}{\ell_D} (L_0 + \tilde{L}_0 - \frac{c+\bar{c}}{24})\,,
\end{equation*}
where $\ell_D$ is the circumference of the region $D$.
Thus, $K_D$ only generates non-trivial modular flow along $\partial D$, where the excitation moves with the speed $v=\beta$. In the case with $U(1)$ symmetry, the CFT is augmented by a holomorphic and anti-holomorphic current $J,\tilde J$ at level $k_L, k_R$. The Hall conductance is $\sigma_{xy} = (k_L-k_R)/2\pi$.

It is more instructive to consider the \emph{$U(1)$ defect operator} $e^{\i\mu Q_{D}}$ and obtain $Q_D^2$ via Taylor expansion in $\mu$.
This operator creates a line defect along the boundary of $D$, and thus its expectation value $Z_{D}(\mu) := \braket{\psi|e^{\i\mu Q_{D}}|\psi}$ satisfies an area-law
\begin{equation}
\label{eq:area law}
    \ln Z_{D}(\mu) = -\alpha \ell_{D} + \gamma + \cdots
\end{equation}
where $\alpha$ is the line tension, $\gamma$ is universal, and the ellipsis represents terms that vanish in the limit $\ell_D \rightarrow \infty$~\cite{Chen:2022jvu}. Our focus is the area-law coefficient, which is dictated by the charged Cardy formula~\cite{Kraus:2006wn}
\begin{equation}
\label{eq:area law coefficient}
    \alpha = \frac{(k_L+k_R)\mu^2}{4\pi \beta}\,.
\end{equation}
This result is related to the spectral flow of the CFT ground state energy under a $U(1)$ twist~\cite{Schwimmer:1986mf}. 
Its linearity in $k_L+k_R$ is because the levels physically quantify the number of charged modes. The $\mu^2$ dependence will play an important role in our following derivation.

\begin{figure}[!t]
\centering
\includegraphics[width=7cm]{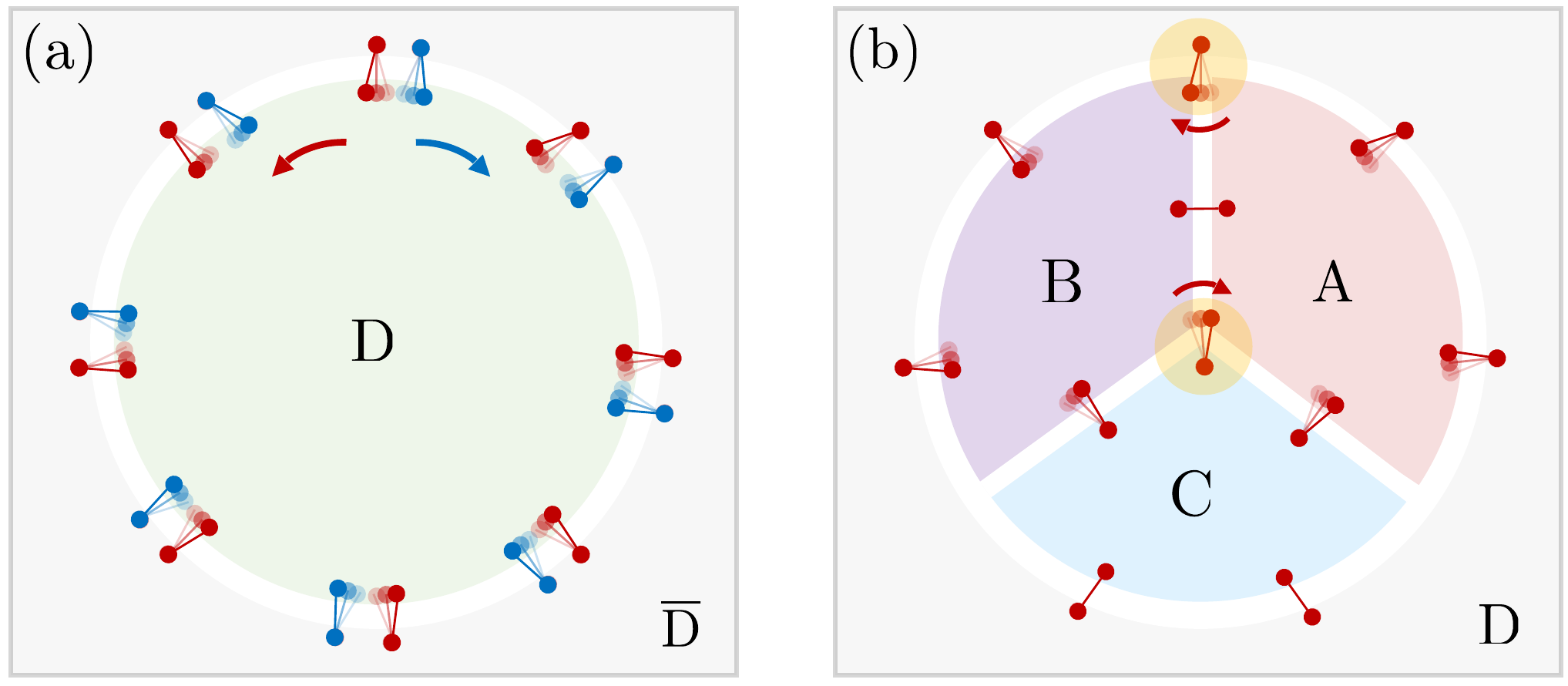}
\caption{(a) Red and blue dots are the chiral and anti-chiral modes, the bonds represents the correlation. The arrows designate their motion under the modular flow $U_D(s)$. (b) The anti-chiral degrees of freedom are suppressed for the clarity of the figure. The arrows designate the motion under the modular flow $U_{AB}(s)$. The yellow shaded disks emphasis modes that make contribution to the change of $Z_{BC}(\mu;s)$ at the linear order in $s$.}
\label{fig:ABC partition}
\end{figure}

We reinterpret the result with a quasi-particle picture. There are chiral and anti-chiral modes sitting near the boundary $\partial D$, the numbers of which are proportional to $k_L$ and $k_R$, respectively. Their correlation across the boundary contributes to the line tension. We can then separate $\alpha$ into two terms that account for the contribution from the two types of modes
\begin{equation}
\label{eq:line tension}
    \alpha_{\text{chiral}} = \frac{k_L \mu^2}{4\pi \beta}\,,\quad
    \alpha_{\text{anti-chiral}} = \frac{k_R \mu^2}{4\pi \beta}\,.
\end{equation}
The idea is sketched in \figref{fig:ABC partition}~(a), where the chiral and anti-chiral modes are the red and blue dots respectively. Under the modular flow generated by $K_D$, the chiral and anti-chiral modes inside the region $D$ move in opposite directions, as shown by the arrows.

Corresponding to \eqnref{eq:sigmaxy formula 1}, we apply this picture to compute the response of $e^{\i\mu Q_{BC}}$ under the modular flow generated by $K_{AB}$
\begin{equation*}
    Z_{BC}(\mu;s):=\braket{\psi|e^{\i sK_{AB}} e^{\i\mu Q_{BC}} e^{-\i sK_{AB}}|\psi}\,.
\end{equation*}
It suffices to focus on the chiral modes, the contribution from the anti-chiral ones is the opposite. 
See \figref{fig:ABC partition}~(b). Only the motion of chiral modes that are near the $ABC$ and $ABD$ triple-contact points can affect $Z_{BC}(\mu;s)$ at the linear order in $s$. Near the $ABC$ triple-contact point, chiral modes in $B$ that are correlated with $C$ will enter $A$ and increase the line tension. The $ABD$ triple-contact point can be analyzed similarly. After summing up the two contributions, we have
\begin{equation*}
    \frac{d}{ds} \ln Z_{BC}(\mu;s) \big|_{s=0} = -2(\alpha_{\text{chiral}} - \alpha_{\text{anti-chiral}}) v
\end{equation*}
where the velocity is $v=\beta$ as introduced before.
By plugging in \eqnref{eq:line tension}, we have
\begin{equation}
\label{eq:2+1d result}
    \frac{d}{ds} \ln Z_{BC}(\mu;s) \big|_{s=0} = -\frac{k_L-k_R}{2\pi} \mu^2 = -\sigma_{xy} \mu^2\,,
\end{equation}
Here the non-universal factor $\beta$ is completely canceled, which yields a universal answer. This is the main result of this work.
Expanding $e^{\i\mu Q_{BC}}$ on the left-hand side above to quadratic order in $\mu$ gives the right-hand side of \eqnref{eq:sigmaxy formula 1} which verifies our identification in \eqnref{eq:conjecture}. 
In \appref{app:anomaly}, we apply the same idea to purely 1+1D CFTs and show that it detects the perturbative chiral anomaly, which is consistent with the spirit of the bulk-edge correspondence.

\begin{figure}[!t]
\centering
\includegraphics[width=8.5cm]{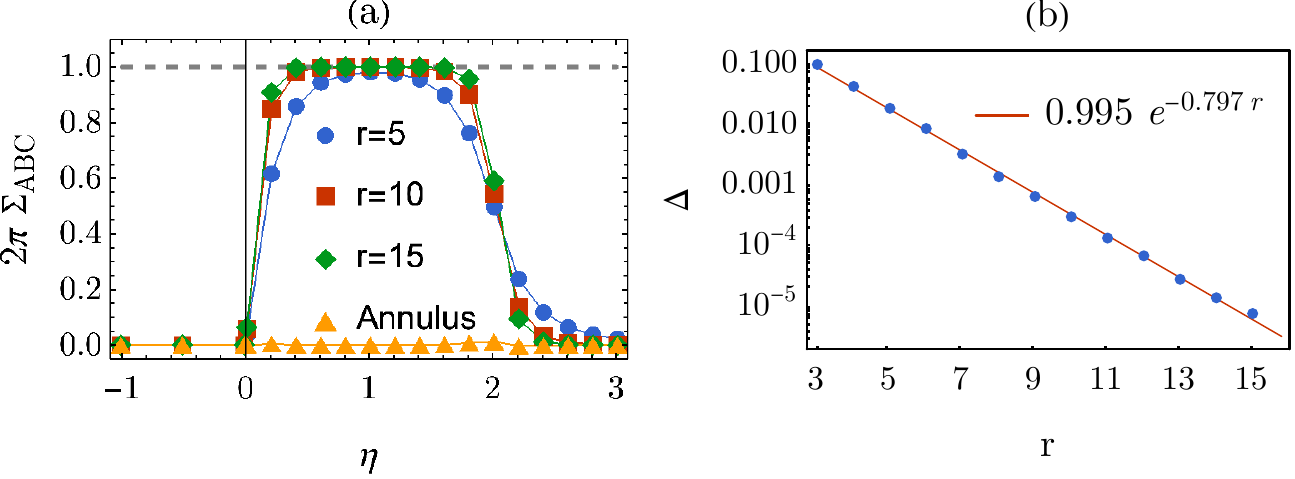}
\caption{(a) $\Sigma(\psi,A,B,C)$ across the phase diagram. Blue, red and green curves correspond to the disk geometry at the corresponding radii $r$. The dashed line is the value of the quantized Hall conductance. (b) Finite size scaling. The blue dots are the data points and the red line is the fitting result. We choose $L_x=L_y=40$. For the annulus geometry, the inner and outer radius is $7$ and $14$. For (b), the inverse correlation length is $\xi^{-1} = 0.398$.}
\label{fig:phase transition}
\end{figure}

\emph{Numerics}--- 
We provide numerical support to our conjecture by simulating free fermion systems. In this section we report on a lattice model and focuses on $\Sigma(\psi;A,B,C)$ in different phases, its finite size scaling and topological rigidity.
\appref{app:details of numerics} contains other aspects as well as results on a continuum model.
In particular, we numerically check the remarkable $\mu^2$ dependence in the area-law coefficient \eqnref{eq:area law coefficient} in lattice systems and find excellent agreement.

We simulate the $\pi$-flux model on a square lattice. The Hamiltonian, $H = \sum_{ij}t_{ij} c_i^\dag c_j$, consists of nearest-neighbor hopping terms that implement $\pi$-flux per plaquette and next-nearest-neighbor hoppings that open up a gap and give rise to various phases. 
We choose a 1-dimensional path inside the entire phase diagram that is parameterized by $\eta\in\bbR$. 
The system is time-reversal symmetric for $\eta\leq 0$ and breaks the symmetry for $\eta>0$.
There is a topological phase at $0<\eta<2$ with a Chern number $C=1$ and two trivial phases elsewhere. In the numerics, we also add weak on-site disorders to make the situation more general.

The result of $\Sigma(\psi;A,B,C)$ across the phase diagram is shown in \figref{fig:phase transition}~(a).
When $A,B,C$ form a complete disk, as required by the definition of the formula, we obtain the correct Hall conductance of each phase. 
Our formula indeed vanishes identically in the time-reversal symmetric region $\eta<0$.
Near the two phase boundaries, the curve becomes steeper for larger subsystem sizes.
In the topological phase, we perform a more detailed finite size scaling. Let $\Delta := 1- 2\pi \Sigma(\psi;A,B,C)$ be the difference between the theoretical and numerical value, and the result is shown in \figref{fig:phase transition}~(b). The deviation decays exponentially with the linear size of the subsystem and the decaying exponent is twice of the correlation length.
As a sanity check, if $A,B,C$ form an annulus (replace a disk neighborhood of the $ABC$ triple-contact point by region $D$) we find a null result. This is consistent with our quasi-particle picture.

\begin{figure}[!t]
\centering
\includegraphics[width=8.5cm]{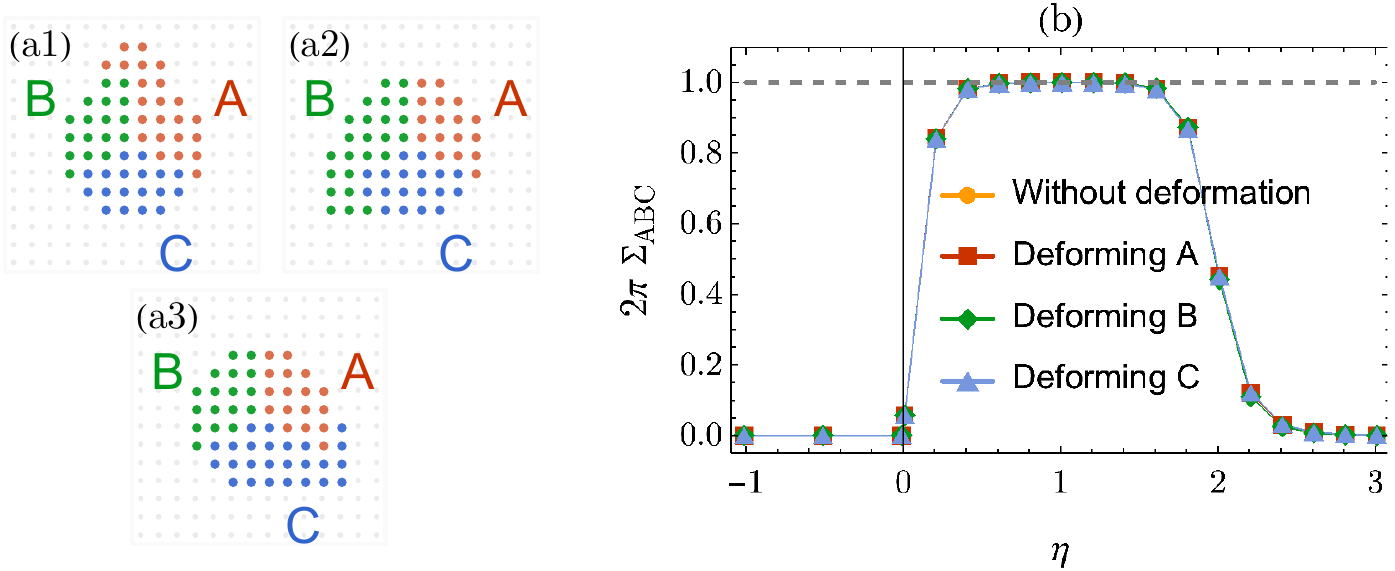}
\caption{Topological rigidity against deformation near the triple-contact points. (a1)-(a3) Attach a blob to region $A,B$ and $C$ near the triple-contact point, respectively. (b) Evaluate $\Sigma(\psi,A,B,C)$ for the deformed geometry across the phase diagram. We choose $L_x=L_y=30$ and the radius of the disk $ABC$ to be $r=10$. The blob has a linear size of $4$.}
\label{fig:topological rigidity}
\end{figure}

We also provide numerical evidence on the topological rigidity as independent support from our analytical argument. For example, we deform the boundaries near the triple-contact points, depicted in \figref{fig:topological rigidity}~(a1)-(a3), and compute $\Sigma(\psi)$ for the deformed geometry. 
As shown in \figref{fig:topological rigidity}~(b), its value remain the same across the entire phase diagram even though the added blobs are not much larger than the correlation length.
The results are the same for other deformations. 

\emph{Discussion}---The present results can be understood as linear response extensions that go beyond `static' entanglement features by utilizing the dynamics generated by the modular (or entanglement) Hamiltonian. Here, the charged Cardy formula evaluates the `static' features of the $U(1)$ twist operator $e^{i\mu Q}$~\cite{Chen:2022jvu}, and its linear response under the modular flow is shown to give the Hall conductance.
This perspective also suggests new potential connections between conventional responses and quantum entanglement. For instance, in free fermion systems with single-charge fermion, we may anticipate an entanglement version of the Wiedemann–Franz (WF) law, which indeed can be obtained by combining the result in \cite{Kim:2021gjx} and our formula:
\begin{equation}
\braket{[K_{AB}, K_{BC} - \frac{\pi^2}{3} Q_{BC}^2]}_\psi = 0\,.
\end{equation}
Understanding the violation of this entanglement WF law and finding other such relations can shed new light on the study of entanglement of many-body systems.

Entanglement entropy suffers from UV divergence in generic Lorentz invariant quantum field theories (QFTs). Distilling universal information from entanglement requires one to carefully define  entanglement in QFTs and further distill the universal part of it.
Our formula, shown to be universal, is an interesting quantity to compute especially in topological quantum field theories. 
Formulating the calculation in a meaningful way requires a better understanding of multipartite entanglement in QFTs, which is an interesting and nontrivial question in its own right~\footnote{We thank Daniel Jafferis and Xi Yin for discussions on this point.}.

\bigskip

\textit{Acknowledgements}---We thank John Cardy, Yiming Chen, Meng Cheng, Joel Moore, Huajia Wang, Xi Yin, Carolyn Zhang, and especially Yingfei Gu, Daniel Jafferis, Ari Turner and Michael Zaletel for many helpful discussions and comments.
RF and AV are supported by a Simons Investigator award (AV) and by the Simons Collaboration on Ultra-Quantum Matter, which is a grant from the Simons Foundation (651440, AV).
RF is supported by the DARPA DRINQS program (award D18AC00033).
This work is funded in part by a QuantEmX grant from ICAM and the Gordon and Betty Moore Foundation through Grant GBMF9616 to Ruihua Fan.
RS acknowledges support by the U.S. Department of Energy, Office of Science, Office of Advanced Scientific Computing Research, Department of
Energy Computational Science Graduate Fellowship under Award Number DESC0022158. 

\bibliography{ref.bib}

\begin{thebibliography}{42}%
\makeatletter
\providecommand \@ifxundefined [1]{%
 \@ifx{#1\undefined}
}%
\providecommand \@ifnum [1]{%
 \ifnum #1\expandafter \@firstoftwo
 \else \expandafter \@secondoftwo
 \fi
}%
\providecommand \@ifx [1]{%
 \ifx #1\expandafter \@firstoftwo
 \else \expandafter \@secondoftwo
 \fi
}%
\providecommand \natexlab [1]{#1}%
\providecommand \enquote  [1]{``#1''}%
\providecommand \bibnamefont  [1]{#1}%
\providecommand \bibfnamefont [1]{#1}%
\providecommand \citenamefont [1]{#1}%
\providecommand \href@noop [0]{\@secondoftwo}%
\providecommand \href [0]{\begingroup \@sanitize@url \@href}%
\providecommand \@href[1]{\@@startlink{#1}\@@href}%
\providecommand \@@href[1]{\endgroup#1\@@endlink}%
\providecommand \@sanitize@url [0]{\catcode `\\12\catcode `\$12\catcode
  `\&12\catcode `\#12\catcode `\^12\catcode `\_12\catcode `\%12\relax}%
\providecommand \@@startlink[1]{}%
\providecommand \@@endlink[0]{}%
\providecommand \url  [0]{\begingroup\@sanitize@url \@url }%
\providecommand \@url [1]{\endgroup\@href {#1}{\urlprefix }}%
\providecommand \urlprefix  [0]{URL }%
\providecommand \Eprint [0]{\href }%
\providecommand \doibase [0]{https://doi.org/}%
\providecommand \selectlanguage [0]{\@gobble}%
\providecommand \bibinfo  [0]{\@secondoftwo}%
\providecommand \bibfield  [0]{\@secondoftwo}%
\providecommand \translation [1]{[#1]}%
\providecommand \BibitemOpen [0]{}%
\providecommand \bibitemStop [0]{}%
\providecommand \bibitemNoStop [0]{.\EOS\space}%
\providecommand \EOS [0]{\spacefactor3000\relax}%
\providecommand \BibitemShut  [1]{\csname bibitem#1\endcsname}%
\let\auto@bib@innerbib\@empty
\bibitem [{\citenamefont {Qi}\ and\ \citenamefont {Zhang}(2011)}]{Qi:2010qag}%
  \BibitemOpen
  \bibfield  {author} {\bibinfo {author} {\bibfnamefont {X.~L.}\ \bibnamefont
  {Qi}}\ and\ \bibinfo {author} {\bibfnamefont {S.~C.}\ \bibnamefont {Zhang}},\
  }\bibfield  {title} {\bibinfo {title} {{Topological insulators and
  superconductors}},\ }\href {https://doi.org/10.1103/RevModPhys.83.1057}
  {\bibfield  {journal} {\bibinfo  {journal} {Rev. Mod. Phys.}\ }\textbf
  {\bibinfo {volume} {83}},\ \bibinfo {pages} {1057} (\bibinfo {year}
  {2011})},\ \Eprint {https://arxiv.org/abs/1008.2026} {arXiv:1008.2026
  [cond-mat.mes-hall]} \BibitemShut {NoStop}%
\bibitem [{\citenamefont {Chiu}\ \emph {et~al.}(2016)\citenamefont {Chiu},
  \citenamefont {Teo}, \citenamefont {Schnyder},\ and\ \citenamefont
  {Ryu}}]{Chiu:2015mfr}%
  \BibitemOpen
  \bibfield  {author} {\bibinfo {author} {\bibfnamefont {C.-K.}\ \bibnamefont
  {Chiu}}, \bibinfo {author} {\bibfnamefont {J.~C.~Y.}\ \bibnamefont {Teo}},
  \bibinfo {author} {\bibfnamefont {A.~P.}\ \bibnamefont {Schnyder}},\ and\
  \bibinfo {author} {\bibfnamefont {S.}~\bibnamefont {Ryu}},\ }\bibfield
  {title} {\bibinfo {title} {{Classification of topological quantum matter with
  symmetries}},\ }\href {https://doi.org/10.1103/RevModPhys.88.035005}
  {\bibfield  {journal} {\bibinfo  {journal} {Rev. Mod. Phys.}\ }\textbf
  {\bibinfo {volume} {88}},\ \bibinfo {pages} {035005} (\bibinfo {year}
  {2016})},\ \Eprint {https://arxiv.org/abs/1505.03535} {arXiv:1505.03535
  [cond-mat.mes-hall]} \BibitemShut {NoStop}%
\bibitem [{\citenamefont {{Wen}}(2017)}]{xiaogangReview}%
  \BibitemOpen
  \bibfield  {author} {\bibinfo {author} {\bibfnamefont {X.-G.}\ \bibnamefont
  {{Wen}}},\ }\bibfield  {title} {\bibinfo {title} {{Colloquium: Zoo of
  quantum-topological phases of matter}},\ }\href
  {https://doi.org/10.1103/RevModPhys.89.041004} {\bibfield  {journal}
  {\bibinfo  {journal} {Reviews of Modern Physics}\ }\textbf {\bibinfo {volume}
  {89}},\ \bibinfo {eid} {041004} (\bibinfo {year} {2017})},\ \Eprint
  {https://arxiv.org/abs/1610.03911} {arXiv:1610.03911 [cond-mat.str-el]}
  \BibitemShut {NoStop}%
\bibitem [{\citenamefont {{Zeng}}\ \emph {et~al.}(2015)\citenamefont {{Zeng}},
  \citenamefont {{Chen}}, \citenamefont {{Zhou}},\ and\ \citenamefont
  {{Wen}}}]{XiaoGangQIBook}%
  \BibitemOpen
  \bibfield  {author} {\bibinfo {author} {\bibfnamefont {B.}~\bibnamefont
  {{Zeng}}}, \bibinfo {author} {\bibfnamefont {X.}~\bibnamefont {{Chen}}},
  \bibinfo {author} {\bibfnamefont {D.-L.}\ \bibnamefont {{Zhou}}},\ and\
  \bibinfo {author} {\bibfnamefont {X.-G.}\ \bibnamefont {{Wen}}},\ }\bibfield
  {title} {\bibinfo {title} {{Quantum Information Meets Quantum Matter -- From
  Quantum Entanglement to Topological Phase in Many-Body Systems}},\
  }\href@noop {} {\bibfield  {journal} {\bibinfo  {journal} {arXiv e-prints}\
  ,\ \bibinfo {eid} {arXiv:1508.02595}} (\bibinfo {year} {2015})},\ \Eprint
  {https://arxiv.org/abs/1508.02595} {arXiv:1508.02595 [cond-mat.str-el]}
  \BibitemShut {NoStop}%
\bibitem [{\citenamefont {Hamma}\ \emph {et~al.}(2005)\citenamefont {Hamma},
  \citenamefont {Ionicioiu},\ and\ \citenamefont {Zanardi}}]{Hamma:2004vdz}%
  \BibitemOpen
  \bibfield  {author} {\bibinfo {author} {\bibfnamefont {A.}~\bibnamefont
  {Hamma}}, \bibinfo {author} {\bibfnamefont {R.}~\bibnamefont {Ionicioiu}},\
  and\ \bibinfo {author} {\bibfnamefont {P.}~\bibnamefont {Zanardi}},\
  }\bibfield  {title} {\bibinfo {title} {{Bipartite entanglement and entropic
  boundary law in lattice spin systems}},\ }\href
  {https://doi.org/10.1103/PhysRevA.71.022315} {\bibfield  {journal} {\bibinfo
  {journal} {Phys. Rev. A}\ }\textbf {\bibinfo {volume} {71}},\ \bibinfo
  {pages} {022315} (\bibinfo {year} {2005})},\ \Eprint
  {https://arxiv.org/abs/quant-ph/0409073} {arXiv:quant-ph/0409073}
  \BibitemShut {NoStop}%
\bibitem [{\citenamefont {{Levin}}\ and\ \citenamefont
  {{Wen}}(2006)}]{LevinWen2005}%
  \BibitemOpen
  \bibfield  {author} {\bibinfo {author} {\bibfnamefont {M.}~\bibnamefont
  {{Levin}}}\ and\ \bibinfo {author} {\bibfnamefont {X.-G.}\ \bibnamefont
  {{Wen}}},\ }\bibfield  {title} {\bibinfo {title} {{Detecting Topological
  Order in a Ground State Wave Function}},\ }\href
  {https://doi.org/10.1103/PhysRevLett.96.110405} {\bibfield  {journal}
  {\bibinfo  {journal} {prl}\ }\textbf {\bibinfo {volume} {96}},\ \bibinfo
  {eid} {110405} (\bibinfo {year} {2006})},\ \Eprint
  {https://arxiv.org/abs/cond-mat/0510613} {arXiv:cond-mat/0510613
  [cond-mat.str-el]} \BibitemShut {NoStop}%
\bibitem [{\citenamefont {Kitaev}\ and\ \citenamefont
  {Preskill}(2006)}]{Kitaev:2005dm}%
  \BibitemOpen
  \bibfield  {author} {\bibinfo {author} {\bibfnamefont {A.}~\bibnamefont
  {Kitaev}}\ and\ \bibinfo {author} {\bibfnamefont {J.}~\bibnamefont
  {Preskill}},\ }\bibfield  {title} {\bibinfo {title} {{Topological
  entanglement entropy}},\ }\href
  {https://doi.org/10.1103/PhysRevLett.96.110404} {\bibfield  {journal}
  {\bibinfo  {journal} {Phys. Rev. Lett.}\ }\textbf {\bibinfo {volume} {96}},\
  \bibinfo {pages} {110404} (\bibinfo {year} {2006})},\ \Eprint
  {https://arxiv.org/abs/hep-th/0510092} {arXiv:hep-th/0510092} \BibitemShut
  {NoStop}%
\bibitem [{\citenamefont {Zhang}\ \emph {et~al.}(2012)\citenamefont {Zhang},
  \citenamefont {Grover}, \citenamefont {Turner}, \citenamefont {Oshikawa},\
  and\ \citenamefont {Vishwanath}}]{Zhang:2011jd}%
  \BibitemOpen
  \bibfield  {author} {\bibinfo {author} {\bibfnamefont {Y.}~\bibnamefont
  {Zhang}}, \bibinfo {author} {\bibfnamefont {T.}~\bibnamefont {Grover}},
  \bibinfo {author} {\bibfnamefont {A.}~\bibnamefont {Turner}}, \bibinfo
  {author} {\bibfnamefont {M.}~\bibnamefont {Oshikawa}},\ and\ \bibinfo
  {author} {\bibfnamefont {A.}~\bibnamefont {Vishwanath}},\ }\bibfield  {title}
  {\bibinfo {title} {{Quasi-particle Statistics and Braiding from Ground State
  Entanglement}},\ }\href {https://doi.org/10.1103/PhysRevB.85.235151}
  {\bibfield  {journal} {\bibinfo  {journal} {Phys. Rev. B}\ }\textbf {\bibinfo
  {volume} {85}},\ \bibinfo {pages} {235151} (\bibinfo {year} {2012})},\
  \Eprint {https://arxiv.org/abs/1111.2342} {arXiv:1111.2342 [cond-mat.str-el]}
  \BibitemShut {NoStop}%
\bibitem [{\citenamefont {Cincio}\ and\ \citenamefont
  {Vidal}(2013)}]{VidalCincio}%
  \BibitemOpen
  \bibfield  {author} {\bibinfo {author} {\bibfnamefont {L.}~\bibnamefont
  {Cincio}}\ and\ \bibinfo {author} {\bibfnamefont {G.}~\bibnamefont {Vidal}},\
  }\bibfield  {title} {\bibinfo {title} {Characterizing topological order by
  studying the ground states on an infinite cylinder},\ }\href
  {https://doi.org/10.1103/PhysRevLett.110.067208} {\bibfield  {journal}
  {\bibinfo  {journal} {Phys. Rev. Lett.}\ }\textbf {\bibinfo {volume} {110}},\
  \bibinfo {pages} {067208} (\bibinfo {year} {2013})}\BibitemShut {NoStop}%
\bibitem [{\citenamefont {{Li}}\ and\ \citenamefont
  {{Haldane}}(2008)}]{HaldaneLi}%
  \BibitemOpen
  \bibfield  {author} {\bibinfo {author} {\bibfnamefont {H.}~\bibnamefont
  {{Li}}}\ and\ \bibinfo {author} {\bibfnamefont {F.~D.~M.}\ \bibnamefont
  {{Haldane}}},\ }\bibfield  {title} {\bibinfo {title} {{Entanglement Spectrum
  as a Generalization of Entanglement Entropy: Identification of Topological
  Order in Non-Abelian Fractional Quantum Hall Effect States}},\ }\bibfield
  {journal} {\bibinfo  {journal} {prl}\ }\textbf {\bibinfo {volume} {101}},\
  \href {https://doi.org/10.1103/PhysRevLett.101.010504}
  {10.1103/PhysRevLett.101.010504} (\bibinfo {year} {2008}),\ \Eprint
  {https://arxiv.org/abs/0805.0332} {arXiv:0805.0332 [cond-mat.mes-hall]}
  \BibitemShut {NoStop}%
\bibitem [{\citenamefont {{Fidkowski}}(2010)}]{Fidkowski2009}%
  \BibitemOpen
  \bibfield  {author} {\bibinfo {author} {\bibfnamefont {L.}~\bibnamefont
  {{Fidkowski}}},\ }\bibfield  {title} {\bibinfo {title} {{Entanglement
  Spectrum of Topological Insulators and Superconductors}},\ }\href
  {https://doi.org/10.1103/PhysRevLett.104.130502} {\bibfield  {journal}
  {\bibinfo  {journal} {\prl}\ }\textbf {\bibinfo {volume} {104}},\ \bibinfo
  {eid} {130502} (\bibinfo {year} {2010})},\ \Eprint
  {https://arxiv.org/abs/0909.2654} {arXiv:0909.2654 [cond-mat.str-el]}
  \BibitemShut {NoStop}%
\bibitem [{\citenamefont {{Turner}}\ \emph {et~al.}(2009)\citenamefont
  {{Turner}}, \citenamefont {{Zhang}},\ and\ \citenamefont
  {{Vishwanath}}}]{Ari2009}%
  \BibitemOpen
  \bibfield  {author} {\bibinfo {author} {\bibfnamefont {A.~M.}\ \bibnamefont
  {{Turner}}}, \bibinfo {author} {\bibfnamefont {Y.}~\bibnamefont {{Zhang}}},\
  and\ \bibinfo {author} {\bibfnamefont {A.}~\bibnamefont {{Vishwanath}}},\
  }\bibfield  {title} {\bibinfo {title} {{Band Topology of Insulators via the
  Entanglement Spectrum}},\ }\href@noop {} {\bibfield  {journal} {\bibinfo
  {journal} {arXiv e-prints}\ ,\ \bibinfo {eid} {arXiv:0909.3119}} (\bibinfo
  {year} {2009})},\ \Eprint {https://arxiv.org/abs/0909.3119} {arXiv:0909.3119
  [cond-mat.str-el]} \BibitemShut {NoStop}%
\bibitem [{\citenamefont {{Pollmann}}\ \emph {et~al.}(2010)\citenamefont
  {{Pollmann}}, \citenamefont {{Turner}}, \citenamefont {{Berg}},\ and\
  \citenamefont {{Oshikawa}}}]{PollmannEntSPT}%
  \BibitemOpen
  \bibfield  {author} {\bibinfo {author} {\bibfnamefont {F.}~\bibnamefont
  {{Pollmann}}}, \bibinfo {author} {\bibfnamefont {A.~M.}\ \bibnamefont
  {{Turner}}}, \bibinfo {author} {\bibfnamefont {E.}~\bibnamefont {{Berg}}},\
  and\ \bibinfo {author} {\bibfnamefont {M.}~\bibnamefont {{Oshikawa}}},\
  }\bibfield  {title} {\bibinfo {title} {{Entanglement spectrum of a
  topological phase in one dimension}},\ }\bibfield  {journal} {\bibinfo
  {journal} {prb}\ }\textbf {\bibinfo {volume} {81}},\ \href
  {https://doi.org/10.1103/PhysRevB.81.064439} {10.1103/PhysRevB.81.064439}
  (\bibinfo {year} {2010}),\ \Eprint {https://arxiv.org/abs/0910.1811}
  {arXiv:0910.1811 [cond-mat.str-el]} \BibitemShut {NoStop}%
\bibitem [{\citenamefont {Swingle}\ and\ \citenamefont
  {Senthil}(2012)}]{Swingle:2011hu}%
  \BibitemOpen
  \bibfield  {author} {\bibinfo {author} {\bibfnamefont {B.}~\bibnamefont
  {Swingle}}\ and\ \bibinfo {author} {\bibfnamefont {T.}~\bibnamefont
  {Senthil}},\ }\bibfield  {title} {\bibinfo {title} {{A Geometric proof of the
  equality between entanglement and edge spectra}},\ }\href
  {https://doi.org/10.1103/PhysRevB.86.045117} {\bibfield  {journal} {\bibinfo
  {journal} {Phys. Rev. B}\ }\textbf {\bibinfo {volume} {86}},\ \bibinfo
  {pages} {045117} (\bibinfo {year} {2012})},\ \Eprint
  {https://arxiv.org/abs/1109.1283} {arXiv:1109.1283 [cond-mat.str-el]}
  \BibitemShut {NoStop}%
\bibitem [{\citenamefont {{Chandran}}\ \emph {et~al.}(2011)\citenamefont
  {{Chandran}}, \citenamefont {{Hermanns}}, \citenamefont {{Regnault}},\ and\
  \citenamefont {{Bernevig}}}]{Chandran2011}%
  \BibitemOpen
  \bibfield  {author} {\bibinfo {author} {\bibfnamefont {A.}~\bibnamefont
  {{Chandran}}}, \bibinfo {author} {\bibfnamefont {M.}~\bibnamefont
  {{Hermanns}}}, \bibinfo {author} {\bibfnamefont {N.}~\bibnamefont
  {{Regnault}}},\ and\ \bibinfo {author} {\bibfnamefont {B.~A.}\ \bibnamefont
  {{Bernevig}}},\ }\bibfield  {title} {\bibinfo {title} {{Bulk-edge
  correspondence in entanglement spectra}},\ }\href
  {https://doi.org/10.1103/PhysRevB.84.205136} {\bibfield  {journal} {\bibinfo
  {journal} {\prb}\ }\textbf {\bibinfo {volume} {84}},\ \bibinfo {eid} {205136}
  (\bibinfo {year} {2011})},\ \Eprint {https://arxiv.org/abs/1102.2218}
  {arXiv:1102.2218 [cond-mat.str-el]} \BibitemShut {NoStop}%
\bibitem [{\citenamefont {{Qi}}\ \emph {et~al.}(2012)\citenamefont {{Qi}},
  \citenamefont {{Katsura}},\ and\ \citenamefont {{Ludwig}}}]{QiLudwig}%
  \BibitemOpen
  \bibfield  {author} {\bibinfo {author} {\bibfnamefont {X.-L.}\ \bibnamefont
  {{Qi}}}, \bibinfo {author} {\bibfnamefont {H.}~\bibnamefont {{Katsura}}},\
  and\ \bibinfo {author} {\bibfnamefont {A.~W.~W.}\ \bibnamefont {{Ludwig}}},\
  }\bibfield  {title} {\bibinfo {title} {{General Relationship between the
  Entanglement Spectrum and the Edge State Spectrum of Topological Quantum
  States}},\ }\bibfield  {journal} {\bibinfo  {journal} {prl}\ }\textbf
  {\bibinfo {volume} {108}},\ \href
  {https://doi.org/10.1103/PhysRevLett.108.196402}
  {10.1103/PhysRevLett.108.196402} (\bibinfo {year} {2012}),\ \Eprint
  {https://arxiv.org/abs/1103.5437} {arXiv:1103.5437 [cond-mat.mes-hall]}
  \BibitemShut {NoStop}%
\bibitem [{\citenamefont {Dubail}\ and\ \citenamefont
  {Read}(2015)}]{Dubail:2013pda}%
  \BibitemOpen
  \bibfield  {author} {\bibinfo {author} {\bibfnamefont {J.}~\bibnamefont
  {Dubail}}\ and\ \bibinfo {author} {\bibfnamefont {N.}~\bibnamefont {Read}},\
  }\bibfield  {title} {\bibinfo {title} {{Tensor network trial states for
  chiral topological phases in two dimensions and a no-go theorem in any
  dimension}},\ }\href {https://doi.org/10.1103/PhysRevB.92.205307} {\bibfield
  {journal} {\bibinfo  {journal} {Phys. Rev. B}\ }\textbf {\bibinfo {volume}
  {92}},\ \bibinfo {pages} {205307} (\bibinfo {year} {2015})},\ \Eprint
  {https://arxiv.org/abs/1307.7726} {arXiv:1307.7726 [cond-mat.mes-hall]}
  \BibitemShut {NoStop}%
\bibitem [{\citenamefont {Wahl}\ \emph {et~al.}(2013)\citenamefont {Wahl},
  \citenamefont {Tu}, \citenamefont {Schuch},\ and\ \citenamefont
  {Cirac}}]{Wahl:2013rha}%
  \BibitemOpen
  \bibfield  {author} {\bibinfo {author} {\bibfnamefont {T.~B.}\ \bibnamefont
  {Wahl}}, \bibinfo {author} {\bibfnamefont {H.~H.}\ \bibnamefont {Tu}},
  \bibinfo {author} {\bibfnamefont {N.}~\bibnamefont {Schuch}},\ and\ \bibinfo
  {author} {\bibfnamefont {J.~I.}\ \bibnamefont {Cirac}},\ }\bibfield  {title}
  {\bibinfo {title} {{Projected entangled-pair states can describe chiral
  topological states}},\ }\href
  {https://doi.org/10.1103/PhysRevLett.111.236805} {\bibfield  {journal}
  {\bibinfo  {journal} {Phys. Rev. Lett.}\ }\textbf {\bibinfo {volume} {111}},\
  \bibinfo {pages} {236805} (\bibinfo {year} {2013})},\ \Eprint
  {https://arxiv.org/abs/1308.0316} {arXiv:1308.0316 [cond-mat.str-el]}
  \BibitemShut {NoStop}%
\bibitem [{\citenamefont {{Kapustin}}\ and\ \citenamefont
  {{Fidkowski}}(2019)}]{KapustinFidkowski2018}%
  \BibitemOpen
  \bibfield  {author} {\bibinfo {author} {\bibfnamefont {A.}~\bibnamefont
  {{Kapustin}}}\ and\ \bibinfo {author} {\bibfnamefont {L.}~\bibnamefont
  {{Fidkowski}}},\ }\bibfield  {title} {\bibinfo {title} {{Local Commuting
  Projector Hamiltonians and the Quantum Hall Effect}},\ }\href
  {https://doi.org/10.1007/s00220-019-03444-1} {\bibfield  {journal} {\bibinfo
  {journal} {Communications in Mathematical Physics}\ }\textbf {\bibinfo
  {volume} {373}},\ \bibinfo {pages} {763} (\bibinfo {year} {2019})},\ \Eprint
  {https://arxiv.org/abs/1810.07756} {arXiv:1810.07756 [cond-mat.str-el]}
  \BibitemShut {NoStop}%
\bibitem [{\citenamefont {Siva}\ \emph {et~al.}(2022)\citenamefont {Siva},
  \citenamefont {Zou}, \citenamefont {Soejima}, \citenamefont {Mong},\ and\
  \citenamefont {Zaletel}}]{Siva:2021cgo}%
  \BibitemOpen
  \bibfield  {author} {\bibinfo {author} {\bibfnamefont {K.}~\bibnamefont
  {Siva}}, \bibinfo {author} {\bibfnamefont {Y.}~\bibnamefont {Zou}}, \bibinfo
  {author} {\bibfnamefont {T.}~\bibnamefont {Soejima}}, \bibinfo {author}
  {\bibfnamefont {R.~S.~K.}\ \bibnamefont {Mong}},\ and\ \bibinfo {author}
  {\bibfnamefont {M.~P.}\ \bibnamefont {Zaletel}},\ }\bibfield  {title}
  {\bibinfo {title} {{Universal tripartite entanglement signature of ungappable
  edge states}},\ }\href {https://doi.org/10.1103/PhysRevB.106.L041107}
  {\bibfield  {journal} {\bibinfo  {journal} {Phys. Rev. B}\ }\textbf {\bibinfo
  {volume} {106}},\ \bibinfo {pages} {L041107} (\bibinfo {year} {2022})},\
  \Eprint {https://arxiv.org/abs/2110.11965} {arXiv:2110.11965 [quant-ph]}
  \BibitemShut {NoStop}%
\bibitem [{\citenamefont {Liu}\ \emph {et~al.}(2022)\citenamefont {Liu},
  \citenamefont {Sohal}, \citenamefont {Kudler-Flam},\ and\ \citenamefont
  {Ryu}}]{Liu:2021ctk}%
  \BibitemOpen
  \bibfield  {author} {\bibinfo {author} {\bibfnamefont {Y.}~\bibnamefont
  {Liu}}, \bibinfo {author} {\bibfnamefont {R.}~\bibnamefont {Sohal}}, \bibinfo
  {author} {\bibfnamefont {J.}~\bibnamefont {Kudler-Flam}},\ and\ \bibinfo
  {author} {\bibfnamefont {S.}~\bibnamefont {Ryu}},\ }\bibfield  {title}
  {\bibinfo {title} {{Multipartitioning topological phases by vertex states and
  quantum entanglement}},\ }\href {https://doi.org/10.1103/PhysRevB.105.115107}
  {\bibfield  {journal} {\bibinfo  {journal} {Phys. Rev. B}\ }\textbf {\bibinfo
  {volume} {105}},\ \bibinfo {pages} {115107} (\bibinfo {year} {2022})},\
  \Eprint {https://arxiv.org/abs/2110.11980} {arXiv:2110.11980
  [cond-mat.str-el]} \BibitemShut {NoStop}%
\bibitem [{\citenamefont {Thouless}\ \emph {et~al.}(1982)\citenamefont
  {Thouless}, \citenamefont {Kohmoto}, \citenamefont {Nightingale},\ and\
  \citenamefont {den Nijs}}]{Thouless:1982zz}%
  \BibitemOpen
  \bibfield  {author} {\bibinfo {author} {\bibfnamefont {D.~J.}\ \bibnamefont
  {Thouless}}, \bibinfo {author} {\bibfnamefont {M.}~\bibnamefont {Kohmoto}},
  \bibinfo {author} {\bibfnamefont {M.~P.}\ \bibnamefont {Nightingale}},\ and\
  \bibinfo {author} {\bibfnamefont {M.}~\bibnamefont {den Nijs}},\ }\bibfield
  {title} {\bibinfo {title} {{Quantized Hall Conductance in a Two-Dimensional
  Periodic Potential}},\ }\href {https://doi.org/10.1103/PhysRevLett.49.405}
  {\bibfield  {journal} {\bibinfo  {journal} {Phys. Rev. Lett.}\ }\textbf
  {\bibinfo {volume} {49}},\ \bibinfo {pages} {405} (\bibinfo {year}
  {1982})}\BibitemShut {NoStop}%
\bibitem [{\citenamefont {{Bellissard}}\ \emph {et~al.}(1994)\citenamefont
  {{Bellissard}}, \citenamefont {{van Elst}},\ and\ \citenamefont
  {{Schulz-Baldes}}}]{Bellissard1994}%
  \BibitemOpen
  \bibfield  {author} {\bibinfo {author} {\bibfnamefont {J.}~\bibnamefont
  {{Bellissard}}}, \bibinfo {author} {\bibfnamefont {A.}~\bibnamefont {{van
  Elst}}},\ and\ \bibinfo {author} {\bibfnamefont {H.}~\bibnamefont
  {{Schulz-Baldes}}},\ }\bibfield  {title} {\bibinfo {title} {{The
  noncommutative geometry of the quantum Hall effect}},\ }\href
  {https://doi.org/10.1063/1.530758} {\bibfield  {journal} {\bibinfo  {journal}
  {Journal of Mathematical Physics}\ }\textbf {\bibinfo {volume} {35}},\
  \bibinfo {pages} {5373} (\bibinfo {year} {1994})},\ \Eprint
  {https://arxiv.org/abs/cond-mat/9411052} {arXiv:cond-mat/9411052 [cond-mat]}
  \BibitemShut {NoStop}%
\bibitem [{\citenamefont {Avron}\ \emph {et~al.}(1990)\citenamefont {Avron},
  \citenamefont {Seiler},\ and\ \citenamefont {Simon}}]{Simon1990}%
  \BibitemOpen
  \bibfield  {author} {\bibinfo {author} {\bibfnamefont {J.~E.}\ \bibnamefont
  {Avron}}, \bibinfo {author} {\bibfnamefont {R.}~\bibnamefont {Seiler}},\ and\
  \bibinfo {author} {\bibfnamefont {B.}~\bibnamefont {Simon}},\ }\bibfield
  {title} {\bibinfo {title} {Quantum hall effect and the relative index for
  projections},\ }\href {https://doi.org/10.1103/PhysRevLett.65.2185}
  {\bibfield  {journal} {\bibinfo  {journal} {Phys. Rev. Lett.}\ }\textbf
  {\bibinfo {volume} {65}},\ \bibinfo {pages} {2185} (\bibinfo {year}
  {1990})}\BibitemShut {NoStop}%
\bibitem [{\citenamefont {Kitaev}(2006)}]{Kitaev:2005hzj}%
  \BibitemOpen
  \bibfield  {author} {\bibinfo {author} {\bibfnamefont {A.}~\bibnamefont
  {Kitaev}},\ }\bibfield  {title} {\bibinfo {title} {{Anyons in an exactly
  solved model and beyond}},\ }\href
  {https://doi.org/10.1016/j.aop.2005.10.005} {\bibfield  {journal} {\bibinfo
  {journal} {Annals Phys.}\ }\textbf {\bibinfo {volume} {321}},\ \bibinfo
  {pages} {2} (\bibinfo {year} {2006})},\ \Eprint
  {https://arxiv.org/abs/cond-mat/0506438} {arXiv:cond-mat/0506438}
  \BibitemShut {NoStop}%
\bibitem [{\citenamefont {Bachmann}\ \emph {et~al.}(2021)\citenamefont
  {Bachmann}, \citenamefont {Bols}, \citenamefont {De~Roeck},\ and\
  \citenamefont {Fraas}}]{Bachmann:2020unp}%
  \BibitemOpen
  \bibfield  {author} {\bibinfo {author} {\bibfnamefont {S.}~\bibnamefont
  {Bachmann}}, \bibinfo {author} {\bibfnamefont {A.}~\bibnamefont {Bols}},
  \bibinfo {author} {\bibfnamefont {W.}~\bibnamefont {De~Roeck}},\ and\
  \bibinfo {author} {\bibfnamefont {M.}~\bibnamefont {Fraas}},\ }\bibfield
  {title} {\bibinfo {title} {{Rational indices for quantum ground state
  sectors}},\ }\href {https://doi.org/10.1063/5.0021511} {\bibfield  {journal}
  {\bibinfo  {journal} {J. Math. Phys.}\ }\textbf {\bibinfo {volume} {62}},\
  \bibinfo {pages} {011901} (\bibinfo {year} {2021})},\ \Eprint
  {https://arxiv.org/abs/2001.06458} {arXiv:2001.06458 [math-ph]} \BibitemShut
  {NoStop}%
\bibitem [{\citenamefont {{Kapustin}}\ and\ \citenamefont
  {{Sopenko}}(2020)}]{KapustinHall2020}%
  \BibitemOpen
  \bibfield  {author} {\bibinfo {author} {\bibfnamefont {A.}~\bibnamefont
  {{Kapustin}}}\ and\ \bibinfo {author} {\bibfnamefont {N.}~\bibnamefont
  {{Sopenko}}},\ }\bibfield  {title} {\bibinfo {title} {{Hall conductance and
  the statistics of flux insertions in gapped interacting lattice systems}},\
  }\href {https://doi.org/10.1063/5.0022944} {\bibfield  {journal} {\bibinfo
  {journal} {Journal of Mathematical Physics}\ }\textbf {\bibinfo {volume}
  {61}},\ \bibinfo {eid} {101901} (\bibinfo {year} {2020})},\ \Eprint
  {https://arxiv.org/abs/2006.14151} {arXiv:2006.14151 [math-ph]} \BibitemShut
  {NoStop}%
\bibitem [{\citenamefont {Shiozaki}\ \emph {et~al.}(2018)\citenamefont
  {Shiozaki}, \citenamefont {Shapourian}, \citenamefont {Gomi},\ and\
  \citenamefont {Ryu}}]{Shiozaki:2017ive}%
  \BibitemOpen
  \bibfield  {author} {\bibinfo {author} {\bibfnamefont {K.}~\bibnamefont
  {Shiozaki}}, \bibinfo {author} {\bibfnamefont {H.}~\bibnamefont
  {Shapourian}}, \bibinfo {author} {\bibfnamefont {K.}~\bibnamefont {Gomi}},\
  and\ \bibinfo {author} {\bibfnamefont {S.}~\bibnamefont {Ryu}},\ }\bibfield
  {title} {\bibinfo {title} {{Many-body topological invariants for fermionic
  short-range entangled topological phases protected by antiunitary
  symmetries}},\ }\href {https://doi.org/10.1103/PhysRevB.98.035151} {\bibfield
   {journal} {\bibinfo  {journal} {Phys. Rev. B}\ }\textbf {\bibinfo {volume}
  {98}},\ \bibinfo {pages} {035151} (\bibinfo {year} {2018})},\ \Eprint
  {https://arxiv.org/abs/1710.01886} {arXiv:1710.01886 [cond-mat.str-el]}
  \BibitemShut {NoStop}%
\bibitem [{\citenamefont {Dehghani}\ \emph {et~al.}(2021)\citenamefont
  {Dehghani}, \citenamefont {Cian}, \citenamefont {Hafezi},\ and\ \citenamefont
  {Barkeshli}}]{Dehghani:2020jls}%
  \BibitemOpen
  \bibfield  {author} {\bibinfo {author} {\bibfnamefont {H.}~\bibnamefont
  {Dehghani}}, \bibinfo {author} {\bibfnamefont {Z.-P.}\ \bibnamefont {Cian}},
  \bibinfo {author} {\bibfnamefont {M.}~\bibnamefont {Hafezi}},\ and\ \bibinfo
  {author} {\bibfnamefont {M.}~\bibnamefont {Barkeshli}},\ }\bibfield  {title}
  {\bibinfo {title} {{Extraction of the many-body Chern number from a single
  wave function}},\ }\href {https://doi.org/10.1103/PhysRevB.103.075102}
  {\bibfield  {journal} {\bibinfo  {journal} {Phys. Rev. B}\ }\textbf {\bibinfo
  {volume} {103}},\ \bibinfo {pages} {075102} (\bibinfo {year} {2021})},\
  \Eprint {https://arxiv.org/abs/2005.13677} {arXiv:2005.13677
  [cond-mat.str-el]} \BibitemShut {NoStop}%
\bibitem [{Note1()}]{Note1}%
  \BibitemOpen
  \bibinfo {note} {The close relation between entanglement spectrum (ES) and
  protected edge states suggests a possible strategy based on the ES. However,
  typically, translation symmetry needs to be invoked to fully extract
  information present in the ES.}\BibitemShut {Stop}%
\bibitem [{Note2()}]{Note2}%
  \BibitemOpen
  \bibinfo {note} {One can in principle use the $U(1)$ resolved ES and study
  the spectral flow, which, however, requires a continuous family of states and
  is not convenient to work with.}\BibitemShut {Stop}%
\bibitem [{\citenamefont {Kim}\ \emph {et~al.}(2022)\citenamefont {Kim},
  \citenamefont {Shi}, \citenamefont {Kato},\ and\ \citenamefont
  {Albert}}]{Kim:2021gjx}%
  \BibitemOpen
  \bibfield  {author} {\bibinfo {author} {\bibfnamefont {I.~H.}\ \bibnamefont
  {Kim}}, \bibinfo {author} {\bibfnamefont {B.}~\bibnamefont {Shi}}, \bibinfo
  {author} {\bibfnamefont {K.}~\bibnamefont {Kato}},\ and\ \bibinfo {author}
  {\bibfnamefont {V.~V.}\ \bibnamefont {Albert}},\ }\bibfield  {title}
  {\bibinfo {title} {{Chiral Central Charge from a Single Bulk Wave
  Function}},\ }\href {https://doi.org/10.1103/PhysRevLett.128.176402}
  {\bibfield  {journal} {\bibinfo  {journal} {Phys. Rev. Lett.}\ }\textbf
  {\bibinfo {volume} {128}},\ \bibinfo {pages} {176402} (\bibinfo {year}
  {2022})},\ \Eprint {https://arxiv.org/abs/2110.06932} {arXiv:2110.06932
  [quant-ph]} \BibitemShut {NoStop}%
\bibitem [{\citenamefont {Kim}\ \emph {et~al.}(2021)\citenamefont {Kim},
  \citenamefont {Shi}, \citenamefont {Kato},\ and\ \citenamefont
  {Albert}}]{Kim:2021tse}%
  \BibitemOpen
  \bibfield  {author} {\bibinfo {author} {\bibfnamefont {I.~H.}\ \bibnamefont
  {Kim}}, \bibinfo {author} {\bibfnamefont {B.}~\bibnamefont {Shi}}, \bibinfo
  {author} {\bibfnamefont {K.}~\bibnamefont {Kato}},\ and\ \bibinfo {author}
  {\bibfnamefont {V.~V.}\ \bibnamefont {Albert}},\ }\bibfield  {title}
  {\bibinfo {title} {{Modular commutator in gapped quantum many-body
  systems}},\ }\href@noop {} {\  (\bibinfo {year} {2021})},\ \Eprint
  {https://arxiv.org/abs/2110.10400} {arXiv:2110.10400 [quant-ph]} \BibitemShut
  {NoStop}%
\bibitem [{\citenamefont {Fan}(2022)}]{Fan:2022ukm}%
  \BibitemOpen
  \bibfield  {author} {\bibinfo {author} {\bibfnamefont {R.}~\bibnamefont
  {Fan}},\ }\bibfield  {title} {\bibinfo {title} {{From entanglement generated
  dynamics to the gravitational anomaly and chiral central charge}},\
  }\href@noop {} {\  (\bibinfo {year} {2022})},\ \Eprint
  {https://arxiv.org/abs/2206.02823} {arXiv:2206.02823 [cond-mat.str-el]}
  \BibitemShut {NoStop}%
\bibitem [{\citenamefont {Haag}(1992)}]{Haag:1992hx}%
  \BibitemOpen
  \bibfield  {author} {\bibinfo {author} {\bibfnamefont {R.}~\bibnamefont
  {Haag}},\ }\href@noop {} {\emph {\bibinfo {title} {{Local quantum physics:
  Fields, particles, algebras}}}}\ (\bibinfo {year} {1992})\BibitemShut
  {NoStop}%
\bibitem [{\citenamefont {Kitaev}(2011)}]{KitaevTalk}%
  \BibitemOpen
  \bibfield  {author} {\bibinfo {author} {\bibfnamefont {A.}~\bibnamefont
  {Kitaev}},\ }\bibfield  {title} {\bibinfo {title} {{Toward a topological
  classification of many-body quantum states with short-range entanglement}},\
  }\href@noop {} {\bibfield  {journal} {\bibinfo  {journal} {talk at
  Topological Quantum Computing Workshop}\ } (\bibinfo {year}
  {2011})}\BibitemShut {NoStop}%
\bibitem [{\citenamefont {Chen}\ \emph {et~al.}(2022)\citenamefont {Chen},
  \citenamefont {Tu}, \citenamefont {Meng},\ and\ \citenamefont
  {Cheng}}]{Chen:2022jvu}%
  \BibitemOpen
  \bibfield  {author} {\bibinfo {author} {\bibfnamefont {B.-B.}\ \bibnamefont
  {Chen}}, \bibinfo {author} {\bibfnamefont {H.-H.}\ \bibnamefont {Tu}},
  \bibinfo {author} {\bibfnamefont {Z.~Y.}\ \bibnamefont {Meng}},\ and\
  \bibinfo {author} {\bibfnamefont {M.}~\bibnamefont {Cheng}},\ }\bibfield
  {title} {\bibinfo {title} {{Topological Disorder Parameter}},\ }\href@noop {}
  {\  (\bibinfo {year} {2022})},\ \Eprint {https://arxiv.org/abs/2203.08847}
  {arXiv:2203.08847 [cond-mat.str-el]} \BibitemShut {NoStop}%
\bibitem [{\citenamefont {Kraus}(2008)}]{Kraus:2006wn}%
  \BibitemOpen
  \bibfield  {author} {\bibinfo {author} {\bibfnamefont {P.}~\bibnamefont
  {Kraus}},\ }\bibfield  {title} {\bibinfo {title} {{Lectures on black holes
  and the AdS(3) / CFT(2) correspondence}},\ }\href@noop {} {\bibfield
  {journal} {\bibinfo  {journal} {Lect. Notes Phys.}\ }\textbf {\bibinfo
  {volume} {755}},\ \bibinfo {pages} {193} (\bibinfo {year} {2008})},\ \Eprint
  {https://arxiv.org/abs/hep-th/0609074} {arXiv:hep-th/0609074} \BibitemShut
  {NoStop}%
\bibitem [{\citenamefont {Schwimmer}\ and\ \citenamefont
  {Seiberg}(1987)}]{Schwimmer:1986mf}%
  \BibitemOpen
  \bibfield  {author} {\bibinfo {author} {\bibfnamefont {A.}~\bibnamefont
  {Schwimmer}}\ and\ \bibinfo {author} {\bibfnamefont {N.}~\bibnamefont
  {Seiberg}},\ }\bibfield  {title} {\bibinfo {title} {{Comments on the N=2,
  N=3, N=4 Superconformal Algebras in Two-Dimensions}},\ }\href
  {https://doi.org/10.1016/0370-2693(87)90566-1} {\bibfield  {journal}
  {\bibinfo  {journal} {Phys. Lett. B}\ }\textbf {\bibinfo {volume} {184}},\
  \bibinfo {pages} {191} (\bibinfo {year} {1987})}\BibitemShut {NoStop}%
\bibitem [{Note3()}]{Note3}%
  \BibitemOpen
  \bibinfo {note} {We thank Daniel Jafferis and Xi Yin for discussions on this
  point.}\BibitemShut {Stop}%
\bibitem [{\citenamefont {Ginsparg}(1988)}]{Ginsparg:1988ui}%
  \BibitemOpen
  \bibfield  {author} {\bibinfo {author} {\bibfnamefont {P.~H.}\ \bibnamefont
  {Ginsparg}},\ }\bibfield  {title} {\bibinfo {title} {{APPLIED CONFORMAL FIELD
  THEORY}},\ }in\ \href@noop {} {\emph {\bibinfo {booktitle} {{Les Houches
  Summer School in Theoretical Physics: Fields, Strings, Critical
  Phenomena}}}}\ (\bibinfo {year} {1988})\ \Eprint
  {https://arxiv.org/abs/hep-th/9108028} {arXiv:hep-th/9108028} \BibitemShut
  {NoStop}%
\bibitem [{\citenamefont {Casini}\ \emph {et~al.}(2011)\citenamefont {Casini},
  \citenamefont {Huerta},\ and\ \citenamefont {Myers}}]{Casini:2011kv}%
  \BibitemOpen
  \bibfield  {author} {\bibinfo {author} {\bibfnamefont {H.}~\bibnamefont
  {Casini}}, \bibinfo {author} {\bibfnamefont {M.}~\bibnamefont {Huerta}},\
  and\ \bibinfo {author} {\bibfnamefont {R.~C.}\ \bibnamefont {Myers}},\
  }\bibfield  {title} {\bibinfo {title} {{Towards a derivation of holographic
  entanglement entropy}},\ }\href {https://doi.org/10.1007/JHEP05(2011)036}
  {\bibfield  {journal} {\bibinfo  {journal} {JHEP}\ }\textbf {\bibinfo
  {volume} {05}},\ \bibinfo {pages} {036}},\ \Eprint
  {https://arxiv.org/abs/1102.0440} {arXiv:1102.0440 [hep-th]} \BibitemShut
  {NoStop}%
\end{thebibliography}%

\onecolumngrid
\newpage
\appendix
\setcounter{secnumdepth}{2}

\section{Further details in the proof of general properties}

\subsection{Expectation value of the commutator}
\label{app:first law}

In the main text, we have repeatedly used the following equality in the proof of general properties of the formula $\Sigma(\psi; A,B,C)$
\begin{equation}
\label{eq:commutator}
    \braket{\psi|[K_{AB}, Q_{BC}]|\psi} = 0
\end{equation}
where $A,B,C$ are three disjoint spatial subregions, $K_{AB}$ and $Q_{BC}$ are the modular Hamiltonian and total charge operator for the associated region, respectively.
In this section, we explain \eqnref{eq:commutator} by doing direct manipulations.

Note that $Q_{BC} = Q_B + Q_C$, and we can break the left-hand side of \eqnref{eq:commutator} into two pieces
$$
    \braket{\psi|[K_{AB}, Q_{BC}]|\psi} = \braket{\psi|[K_{AB}, Q_{B}]|\psi} + \braket{\psi|[K_{AB}, Q_{C}]|\psi}\,.
$$
The second piece vanishes because $K_{AB}$ and $Q_C$ are supported on different subregions.
The vanishing of the first term follows from the conversion formula \eqnref{eq:modular duality}.
Let $D$ denote the complement of the subregions $A,B$ and $C$, then we have $K_{AB}\ket{\psi} = K_{CD} \ket{\psi}$ and
$$
    \braket{\psi|[K_{AB}, Q_{B}]|\psi} = \braket{\psi|[K_{CD}, Q_{B}]|\psi}\,,
$$
which obviously vanishes.
More generally, for any operator $O$, we can have
\begin{equation}
	\braket{\psi|[K_{AB},O]|\psi} = 0\,,
\end{equation}
whenever $[K_{CD},O]=0$ even though $[K_{AB},O]$ does not have to be zero.

\subsection{Rigidity}
\label{app:rigidity}

We fill in details on the rigidity of $\Sigma(\psi;A,B,C)$ when the Hamiltonian, or equivalently the state, is locally deformed. We change the state $\psi$ by a local operator $O_x$ supported near $x$, which preserves the $U(1)$ symmetry $[O_x,Q]=0$. For simplicity, we assume that the operator has a zero expectation value and write the deformed state as
\begin{equation}
    \ket\psi \mapsto \ket{\psi_x'} = \frac{1}{\sqrt{Z}}\left(\ket{\psi} + O_x \ket{\psi} \right)
\end{equation}
where $Z = 1 + \braket{O_x^\dag O_x}_\psi$ is the normalization factor of the wavefunction.
The equality we want to show is
\begin{equation}
\label{eq:deform Hamiltonian2}
    \Sigma(\psi;A,B,C) = \Sigma(\psi_x';A,B,C)
\end{equation}
By the clustering property, the correlation functions of local opeartors that are far separated from $x$ are not changed, so are the reduced density matrices and modular Hamiltonians. 

Suppose the operator $O_x$ is inside the bulk of region $C$ and is far from any boundary. The right-hand side is
\begin{equation*}
\begin{aligned}
    \Sigma(\psi_x';A,B,C) =& \frac{-i}{2} \braket{\psi'|[K_{AB}', Q_{BC}^2]|\psi'} \\
    \approx & \frac{-i}{2Z} \braket{\psi|(1+O_x^\dag)[K_{AB}, Q_{BC}^2](1+O_x)|\psi}
\end{aligned}
\end{equation*}
Note that the commutator $[K_{AB}, Q_{BC}^2]$ is supported in region $B$ and is far separated from $x$ so that we can use the clustering property, it follows from which that \eqnref{eq:deform Hamiltonian2} is true. 

When the operator $O_x$ is inside the bulk of region $B$, we can apply the conversion formula for the modular Hamiltonian to first convert $K_{AB}'$ to $K_{CD}$ and repeat the above argument.

\section{Relation to the U(1) chiral anomaly}
\label{app:anomaly}

A nonzero Hall conductance in the (2+1)D bulk corresponds to the perturbative $U(1)$ chiral anomaly on the (1+1)D edge.
It is natural to expect \eqnref{eq:sigmaxy formula 1} to be able to detect the anomaly in a pure (1+1)D CFT. We confirm this intuition by imitating the analysis of the defect operator in the main text.

A generic (1+1)D CFT with a global $U(1)$ symmetry is characterized by a holomorphic and anti-holomorphic current $J, \tilde{J}$. The $U(1)$ charge density $\rho$ is given by $2\pi\rho(x) = J(x) + \tilde J(\bar{x})$.
The two currents are spin-1 primary fields with the following operator-product expansion on the complex $z$-plane~\cite{Ginsparg:1988ui}
\begin{equation}
\begin{aligned}
	J(z) J(0) \sim & \frac{k_L}{z^2}\,,\quad \tilde J(\bar{z}) \tilde J(0) \sim  \frac{k_R}{\bar{z}^2} \,.
\end{aligned}
\end{equation}
where $k_{L}$ and $k_R$ are called the levels. When $k_L \neq k_R$, the system has a perturbative chiral anomaly and cannot be defined on manifolds with a boundary.

Consider a system on a circle of length $L=2\pi$. We assume that the conformal vacuum $\ket{0}$ can be realized as the ground state of a critical lattice system and has a Schmidt decomposition. 
Let $A,B,C$ be three adjacent disjoint intervals shown below.
By conformal invariance, the modular Hamiltonian $K_{AB}$ is a local integral of the stress-energy tensor
\begin{equation}
\label{eq:KD CFT}
\begin{tikzpicture}[scale=0.8,baseline={(current bounding box.center)}]
\draw[gray!50,thick] (0,0) circle (1);
\draw[fill=black] (1,0) circle (0.05) node[right] {\scriptsize$x'=0$ \text{or} $L$};
\draw[fill=black] (-0.174,0.984) circle (0.05) node[above] {\scriptsize$x'=y$};
\draw[fill=black] (-0.985,-0.174) circle (0.05) node[left] {\scriptsize$x'=\ell$};
\draw[fill=black] (0.1736,-0.9848) circle (0.05) node[below] {\scriptsize$x'=x$};
\draw[<->,>=stealth,line width=1.5,Acolor!90!black] (1,0) arc (0:98:1 and 1) node at (0.9,0.9) {$A$};
\draw[<->,>=stealth,line width=1.5,Bcolor!90!black] (-0.19,0.984) arc (100:188:1 and 1) node at (-0.95,0.8) {$B$};
\draw[<->,>=stealth,line width=1.5,Ccolor!90!black] (-0.985,-0.185) arc (190:278:1 and 1) node at (-0.85,-0.95) {$C$};
\end{tikzpicture}\qquad
	K_{AB} = 4\pi \int_{AB} \frac{\sin \frac{\ell-x}{2} \sin \frac{x}{\pi}}{\sin \frac{\ell}{2}}\, T_{00}(x) dx\,,
\end{equation}
where $T_{00} = \frac{1}{2\pi} (T+\tilde T)$ is the energy density ~\cite{Casini:2011kv}. 
Under the modular flow generated by $K_{AB}$, the evolution of $e^{i\mu Q_{BC}}$ can be computed by replacing the defect operator by two twist operators $V_{\pm\mu}$ at the end points, i.e.
\begin{equation}
	e^{i\mu Q_{BC}} \mapsto \delta^{(k_L+k_R)(\frac{\mu}{2\pi})^2} V_\mu(x) V_{-\mu}(y)\,,
\end{equation}
where $\delta$ is a UV length scale, $V_{\pm\mu}$ are primary operators with the conformal dimension $h = \frac{k_L}{2} \left(\frac{\mu}{2\pi} \right)^2$, $\tilde h = \frac{k_R}{2} \left(\frac{\mu}{2\pi} \right)^2$ and are not mutually local with the charged operators~\cite{Schwimmer:1986mf}. 
The linear response is given by
\begin{equation}
\begin{aligned}
	\frac{d}{ds} \log & Z_{BC}(\mu,s) \Big|_{s=0} = -\mu^2\frac{k_L-k_R}{4\pi} \left( 1 - 2 \frac{\sin \frac{x}{2} \sin \frac{\ell-y}{2}}{\sin \frac{\ell}{2} \sin \frac{x-y}{2}} \right) \,.
\end{aligned}
\end{equation}
where the ratio of the sines is the cross ratio of the four points drawn on the circle.
The appearance of $k_L-k_R$ is due to the fact that the chiral and anti-chiral modes move in the opposite direction under the modular flow.
Now, we consider the limit $x\rightarrow 2\pi$ where $ABC$ occupies the entire circle and their union is in the pure state $\rho_{ABC} = \ket{0}\bra{0}$.
On the one hand, the above result reduces to
\begin{equation}
\label{eq:cft result}
	\frac{d}{ds} \log Z_{BC}(\mu;s) \big|_{s=0} = -\frac{k_L-k_R}{4\pi} \mu^2\,.
\end{equation}
On the other hand, we have $K_{AB}\ket{\psi} = K_C\ket{\psi}$. Since $K_{C}$ preserves the total charge in $BC$, $Z_{BC}(\mu;s)$ should not have any time-dependence, which implies $k_L=k_R$ for a well-defined quantum state. An intriguing fact is that \eqnref{eq:2+1d result} differs from \eqnref{eq:cft result} by a factor of $2$, which comes from summing over contribution from the two triple-contact point.

\section{More on the free-fermion numerics}
\label{app:details of numerics}

In this appendix, we provide more details on the free-fermion numerics. 
In addition to the definition of the $\pi$-flux model, we present more numerical results.
We also show results for a continuum model, the integer quantum Hall system at $\nu=1$, which serves as a complement to numerics on the lattice.

\subsection{$\pi$-flux model}

\emph{Definition of the model: } The $\pi$-flux model is defined on a square lattice as follows. There are two sublattices A and B, which are depicted as the red and blue dots in the figure below. 
Each unit cell contains an $A$ and $B$ sublattice site, as is denoted by the circle in the figure.
We use $r$ to label the unit cell, $c_{r,A}$ and $c_{r,B}$ are the fermion annihilation operators. 
The Hamiltonian consists of two parts
\begin{equation}
    H = H_1 + H_2\,.
\end{equation}
The first part contains the nearest-neighbor hopping terms that carry appropriate phases such that fermions can pick up a $\pi$ flux after hopping around each plaquette. Our convention for the phases is analogous to the Landau gauge. As is depicted in the figure, the solid links represent hopping with a positive amplitude $t_1>0$ and the dashed links represent hopping with a negative amplitude $-t_1$, i.e.,
\begin{equation}
\begin{tikzpicture}[scale=0.6, baseline={(current bounding box.center)}]
\foreach \x in {0,1,...,5}{
\draw (\x,-0.3) -- (\x,3.3);
}
\foreach \y in {0,2}{
\draw (-0.3,\y) -- (5.3,\y);
}
\foreach \y in {1,3}{
\draw[dashed] (-0.3,\y) -- (5.3,\y);
}
\foreach \y in {0,2}{
\foreach \x in {0,1,...,5}{
	\filldraw[red!65] (\x,\y) circle (0.05);
}
}
\foreach \y in {1,3}{
\foreach \x in {0,1,...,5}{
	\filldraw[blue!65] (\x,\y) circle (0.05);
}
}
\foreach \y in {0,1,2}{
\foreach \x in {0,1,...,4}{
	\node at (\x+0.5,\y+0.5) {\scriptsize $\pi$};
}
}
\draw (1.2,0.5) arc (0:360:0.2 and 0.8);
\node[red!65,left] at (-0.2,0) {\scriptsize$A$};
\node[blue!65,left] at (-0.2,1) {\scriptsize$B$};
\node[red!65,left] at (-0.2,2) {\scriptsize$A$};
\node[blue!65,left] at (-0.2,3) {\scriptsize$B$};
\end{tikzpicture}\qquad
\begin{aligned}
H_1 = t_1 \sum_{r} & c_{r,A}^\dag c_{r+\hat{x},A} - c_{r,B}^\dag c_{r+\hat{x},B} \\
& + c_{r,A}^\dag c_{r,B} + c_{r,B}^\dag c_{r-\hat{y},A} + h.c.
\end{aligned}
\end{equation}
where $\hat{x}$ and $\hat{y}$ denote the horizontal and vertical directions.
The second part contains a special next-nearest-neighbor hopping term with amplitude $t_2$, which is shown by the arrow below
\begin{equation}
\begin{tikzpicture}[scale=0.6, baseline={(current bounding box.center)}]
\foreach \x in {0,1,...,5}{
\draw (\x,-0.3) -- (\x,3.3);
}
\foreach \y in {0,2}{
\draw (-0.3,\y) -- (5.3,\y);
}
\foreach \y in {1,3}{
\draw[dashed] (-0.3,\y) -- (5.3,\y);
}
\foreach \y in {0,2}{
\foreach \x in {0,1,...,5}{
	\filldraw[red!65] (\x,\y) circle (0.05);
}
}
\foreach \y in {1,3}{
\foreach \x in {0,1,...,5}{
	\filldraw[blue!65] (\x,\y) circle (0.05);
}
}
\foreach \y in {0,2}{
\foreach \x in {0,1,...,4}{
	\draw[mid arrow,black!70] (\x,\y) -- (\x+1,\y+1);
}
}

\node[red!65,left] at (-0.2,0) {\scriptsize$A$};
\node[blue!65,left] at (-0.2,1) {\scriptsize$B$};
\node[red!65,left] at (-0.2,2) {\scriptsize$A$};
\node[blue!65,left] at (-0.2,3) {\scriptsize$B$};
\end{tikzpicture}\qquad
H_2 = \sum_{r} t_2\, c_{r+\hat{x},B}^\dag c_{r,A} + h.c.\,.
\end{equation}
The system is gappless with two Dirac cones if we turn off the next-nearest neighbor hopping, i.e., $t_2=0$. The system is generally gapped when $t_2\neq 0$.

Without loss of generality, we fix $t_1=1$.
When $t_2$ is purely real, the system is time-reversal symmetric and cannot have a nonzero Hall conductance. 
When $t_2$ is purely imaginary, the system breaks the time-reversal symmetry and can become non-trivial. Let $t_2=i|t_2|$, then one can show that the system is a Chern insulator with a Chern number $C=1$ for $0<|t_2|<2$ and a trivial insulator for $|t_2|>2$. In the following, we choose a path in the 2-dimensional complex $t_2$ plane, that is parameterized by $\eta$ such that we have $t_2=|\eta|$ being purely real for $\eta<0$ and $t_2=i|\eta|$ purely imaginary for $\eta>0$. This explains the phase diagram quoted in the main text.

\bigskip

\emph{Modular flow of the defect operator}
We derive our conjecture by studying the linear response of the defect operator $e^{i\mu Q_{BC}}$ under the modular flow generated by $K_{AB}$. 
We did the analysis in the topological limit, i.e., the state is well-captured by a topological quantum field theory, and found that only the $\mu^2$ term appears in the line tension and final expression.
Here, we provide numerical result and provide an independent support to this $\mu^2$ dependence.

\begin{figure}
\centering
\includegraphics[width=14cm]{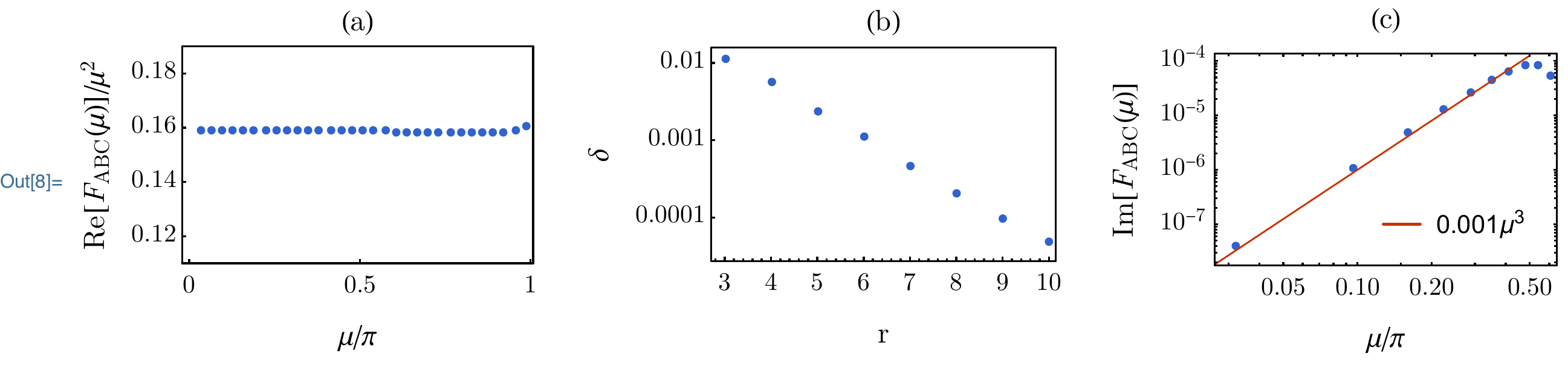}
\caption{Linear response of the defect operator under the modular flow. (a) The real part of $F_{ABC}(\mu)$ is almost quadratic in $\mu$ (b) Finite size scaling of the discrepancy $\delta = \frac{1}{2\pi} - \frac{\Re F_{ABC}(\mu)}{\mu^2}$ (c) The imaginary part of $F_{ABC}(\mu)$ is cubic in $\mu$ for small $\mu$. Figure (a) and (c) uses $L_x=L_y=26$ and $r=7$ as the radius of the subsystem. Figure (b) uses $L_x=L_y=30$, $t_2 = i$ and $\mu=0.1$. Choosing larger values of $\mu$ gives the same result.}
\label{fig:modular flow defect operator}
\end{figure}

For brevity, we denote the linear response of the defect operator by
\begin{equation}
    F_{ABC}(\mu) := \frac{d}{ds} \ln Z_{BC}(\mu;s) \big|_{s=0} = \frac{i\braket{[K_{AB}, e^{i\mu Q_{BC}}]}_\psi}{\braket{e^{i\mu Q_{BC}}}_\psi}
\end{equation}
and numerically study its dependence on $\mu$. The results are shown in \figref{fig:modular flow defect operator}.
\figref{fig:modular flow defect operator}~(a) show that the real part $\Re F_{ABC}(\mu)$ is almost quadratic in $\mu$.
Because the line tension $\alpha$ increases with $\mu$, the absolute value of $\braket{e^{i\mu Q_{BC}}}_\psi$ becomes extremely small near $\mu=\pi$ and causes numerical instability. This may explain the wiggles near $\mu=\pi$. 
We then study the discrepancy between $\Re F_{ABC}(\mu)/\mu^2$ and the Hall conductance and found that it decays exponentially with respect to the subsystem size, which is shown in \figref{fig:modular flow defect operator}~(b).
In the lattice system, $F_{ABC}(\mu)$ also acquires a small imaginary part, which has a cubic dependence in $\mu$ when $\mu$ is small and the result is shown in \figref{fig:modular flow defect operator}~(c). This can be understood via a Taylor expansion of $F_{ABC}(\mu)$ and one can find that the non-vanishing leading term of the imaginary part is cubic in $\mu$.

\emph{Incomplete disk}
The definition of $\Sigma(\psi;A,B,C)$ requires that every three subregions meet once and only once at a point. It is crucial to use this topology in order to get $\sigma_{xy}$, which is manifest from the quasi-particle picture. Here, we present numerical results on what happens when this condition is broken.

\begin{figure}[!t]
\centering
\includegraphics[width=12cm]{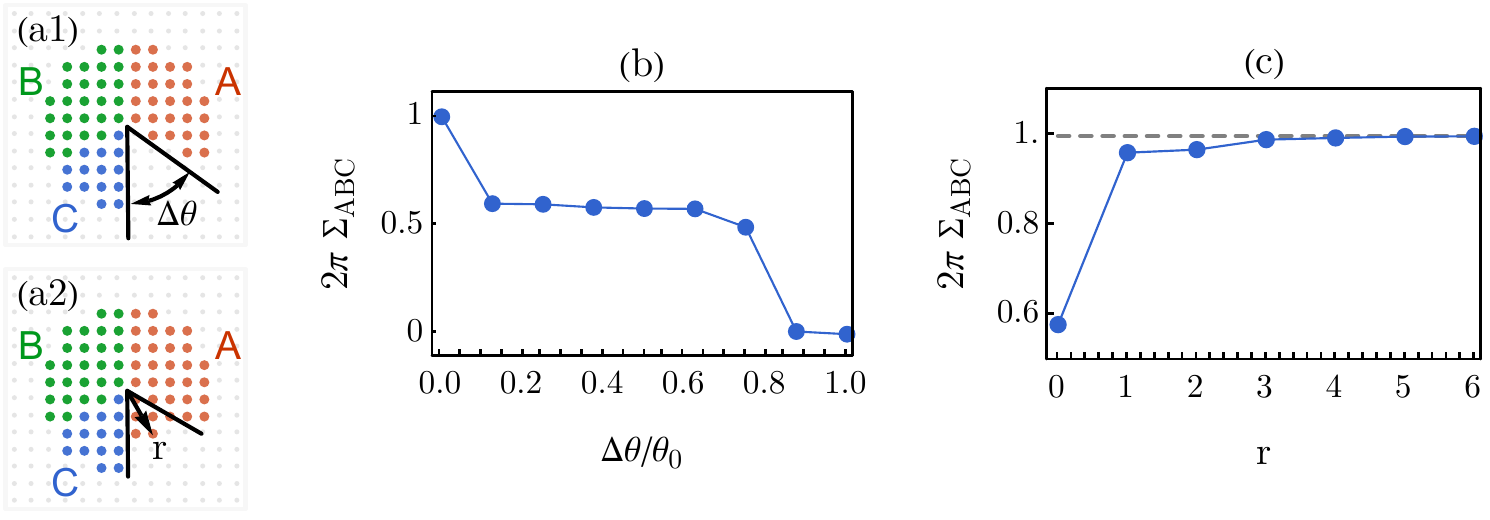}
\caption{Evaluate $\Sigma(\psi;A,B,C)$ for an incomplete disk. (a1) We remove part of region $C$ to have an incomplete disk, the angle of the remove part is $\Delta\theta$. (a2) We add lattice sites that are near the center of the system to region $A$ to remove the quadruple-contact point. (b) $\Sigma(\psi;A,B,C)$ decreases with the angle $\Delta\theta$. (c) $\Sigma(\psi;A,B,C)$ quickly returns to the quantized value as the size of the added region increases. We choose the total system size to be $L_x=L_y=40$ and the radius of the disk $ABC$ to be $R=14$.}
\label{fig:incomplete disk}
\end{figure}

An example is shown in \figref{fig:incomplete disk}~(a1), where region $A,B$ and $C$ are represented by red, green and blue dots, and region $D$ is the rest part.
We remove part of the region $C$ such that it together with $A$ and $B$ only forms an incomplete disk, which replaces the $ACD$ triple-contact point with an $ABCD$ quadruple-contact point.
Let $\theta_0=2\pi/3$ be the angle of subregion $A$ and $B$, $\Delta \theta$ be the angle of the removed part. It is found that $\Sigma(\psi;A,B,C)$ decreases as $\Delta\theta$ increases, which is shown in \figref{fig:incomplete disk}~(b). This behavior is qualitatively consistent with the quasi-particle picture. Namely, near the quadruple contact point, quasi-particles that have correlation between $B$ and $D$ will make opposite contribution compared with ones that correlates $B$ and $C$.

To further demonstrate the importance of having only triple-contact point, we start with an incomplete disk and gradually add lattice sites to region $A$, as is shown in \figref{fig:incomplete disk}~(a2), where the added sites are near the center of the figure. Once the size of the added blob is comparable with the correlation length, it is expected that $\Sigma$ should return to the quantized value, $1$ in this case. This intuition is confirmed by the result shown in \figref{fig:incomplete disk}~(c).

\subsection{Integer quantum Hall system}

In this section, we discuss the integer quantum Hall (IQH) system. 
For simplicity, we will focus on the lowest Landau level and formulate the problem in the continuum as a complementary to the lattice-model discussion.

The system is defined on a finite disk of radius $R$ with a uniform magnetic field $B$ perpendicular to the plane. We choose the symmetric Coulomb gauge $A_x = -By/2$, $A_y = Bx/2$ and write the single-particle Hamiltonian as ($e>0$)
\begin{equation}
H = \frac{1}{2m} (p_x-eA_x)^2 + \frac{1}{2m} (p_y-eA_y)^2\,.
\end{equation}
The system has an intrinsic length scale $\ell_B = \sqrt{1/eB}$, which describes the typical radius of the semi-classical cyclotron motion.
For simplicity, we will set $\ell_B=1$ in the following discussion.

We will focus on the lowest Landau level (LLL), which has $R^2/2$ degenerate states. For $B>0$, these states are given by the following wavefunctions
\begin{equation}
\label{eq:LLL wavefunction}
\varphi_{n}(z) = \frac{1}{\sqrt{2\pi 2^n n!}} z^n e^{-\frac{|z|^2}{4}},\quad n=0,\cdots,\frac{R^2}{2}-1\,,
\end{equation}
where $z=x+iy$ is the holomorphic coordinate. Replacing $z$ with $\bar z = x-iy$ gives the wavefunction for $B<0$.
Strictly speaking, \eqref{eq:LLL wavefunction} are solutions on the infinite plane and cannot be applied to a finite disk. We will choose $R\gg 1$ such that \eqref{eq:LLL wavefunction} is a good approximation to the actual solution away from the boundary.

\begin{figure}[!h]
\centering
\includegraphics[width=12cm]{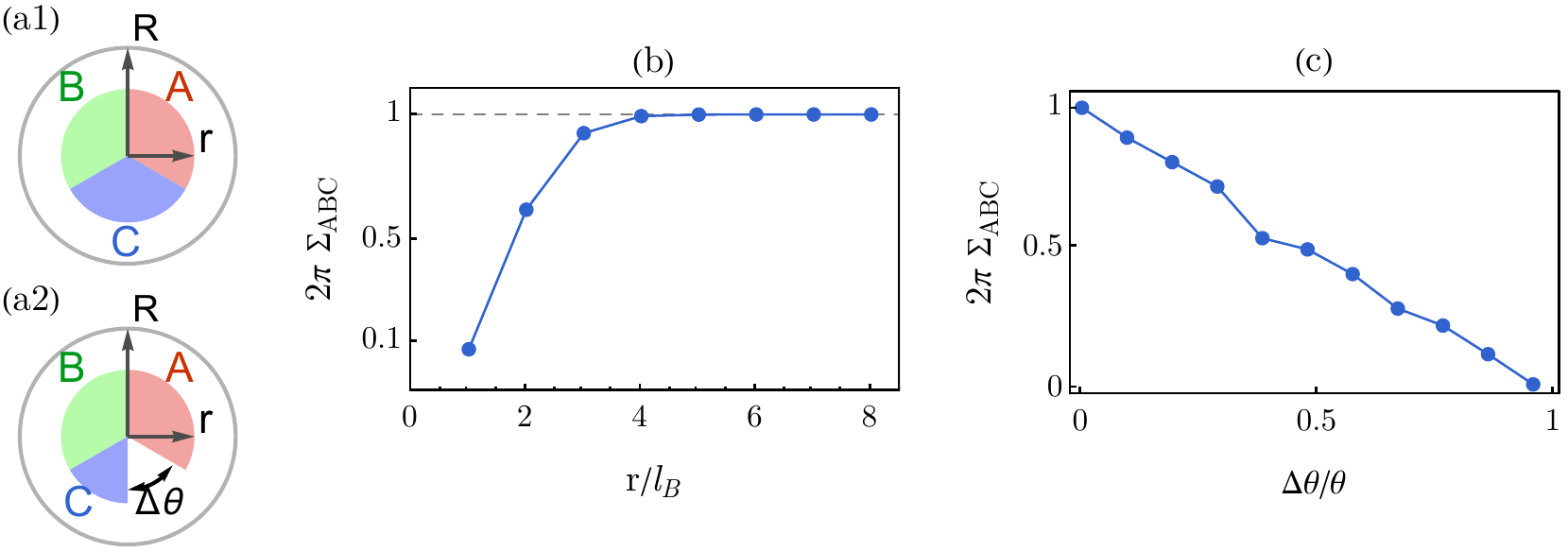}
\caption{Evaluate $\Sigma(\psi;A,B,C)$ for the integer quantum Hall system. (a1) $ABC$ form a complete disk. (a2) $ABC$ form an incomplete disk. (b) $\Sigma(\psi;A,B,C)$ as a function of the radius of the disk. (c) $\Sigma(\psi;A,B,C)$ as a function of the angle of the remove region.}
\label{fig:IQH numerics}
\end{figure}

The results are shown in \figref{fig:IQH numerics}.
We first consider the case where $ABC$ forms a complete disk, as depicted in (a1).
When the radius of the disk becomes much larger than the magnetic length, $2\pi \Sigma(\psi;A,B,C)$ quickly converges to $1$, as shown in (b).
We then consider the case where $ABC$ forms an incomplete disk, as depicted in (a2), i.e., the case with a quadruple-contact point.
As the angle of the removed region increases, $\Sigma(\psi;A,B,C)$ also decreases, which is qualitative the same as what we have observed for the $\pi$-flux model.
One quantitative difference is that $\Sigma(\psi;A,B,C)$ decreases in a smooth way for the IQH system while quite discontinuously for the $\pi$-flux model. Such a difference may be attributed to using a smooth wavefunction in the continuum verses discrete lattice model.

\end{document}